\documentclass[journal]{vgtc}                     

\onlineid{1538}

\vgtccategory{Research}

\title{A Multi-Layer System for Ultra-High-Resolution Static 360-Degree Telepresence}

\author{%
  \authororcid{Jiapeng Chi}{0000-0002-0663-099X},
  \authororcid{Gerd Bruder}{0000-0003-1450-1802},
  \authororcid{Carsten Neumann}{0009-0009-1246-9942},
  \authororcid{Carolina Cruz-Neira}{0000-0002-8499-8909}, and 
  \authororcid{Dirk Reiners}{0000-0002-9344-457X}
}

\authorfooter{
  \item
  The authors are with the VARLab, University of Central Florida,
  Orlando, FL 32816, USA.
  E-mail: \{Jiapeng.Chi, Gerd.Bruder, Carsten.Neumann, carolina, Dirk.Reiners\}@ucf.edu.
}

\abstract{%
  360-degree video telepresence offers strong immersive potential but remains constrained by the limited resolution of current capture and display hardware. Many telepresence installations feature fixed viewpoints and largely static scenes, yet optimization strategies tailored to such setups have received limited attention. We present a multi-layer, ultra-high-resolution system for static 360-degree telepresence that combines an 8K panoramic camera with a rotatable 4K pan-tilt-zoom (PTZ) camera. Our approach builds a three-layer representation: (1) a tile-based ultra-high-resolution panoramic background, generated by offline stitching high-detail 4K PTZ scans onto the base 8K panorama to achieve effective resolution beyond native capture, and represented as a set of spatial tiles; (2) a dynamic update layer that composites foreground motions from the 8K stream via real-time high-resolution background matting; and (3) a region-of-interest 4K layer that streams a real-time PTZ view of the selected region and additionally updates the corresponding background tiles over time. We evaluate the proposed system through comparisons with representative video super-resolution approaches and a user study assessing perceived detail and immersive experience. Our results indicate that tile-based background refinement, together with user-guided updates, provides a practical way to balance panoramic fidelity and interactivity in static 360-degree telepresence.
}


\teaser{
  \centering
  \includegraphics[width=\linewidth]{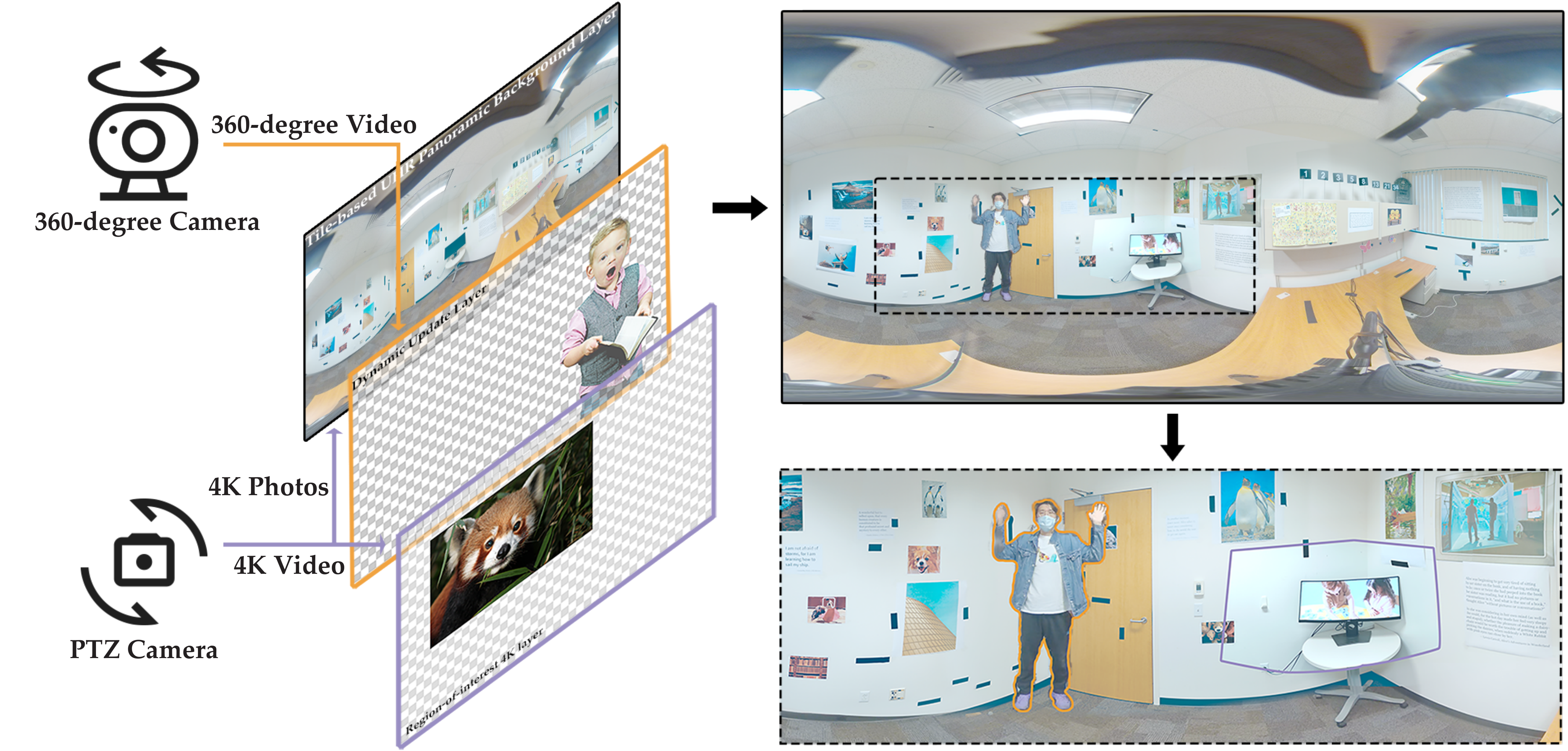}
  \caption{System overview. A 360-degree camera and a PTZ camera jointly produce three layers that are rendered on various display devices. The full scene is represented as an equirectangular panorama, with a highlighted region enlarged below for clarity. Yellow borders indicate the dynamic update layer, purple borders denote the region-of-interest (ROI) 4K layer, and the remaining content corresponds to the tile-based ultra-high-resolution (UHR) panoramic background layer. Images shown in the schematic are example placeholders from the DIV2K dataset \cite{agustsson2017ntire}, intended only to illustrate the role of each layer rather than to depict actual captures or outputs of our system. Icons are adapted from Google Material Icons (Apache License 2.0).}
  \label{fig:teaser}
}

\graphicspath{{figs/}{figures/}{pictures/}{images/}{./}} 

\usepackage{booktabs}                  
\usepackage{lipsum}                    
\usepackage{mwe}                       
\usepackage{ccicons}                   

\usepackage{mathptmx}                  
\usepackage{amsmath}
\usepackage{tabu}
\usepackage{subfigure}
\usepackage{makecell}
\usepackage{booktabs}
\usepackage{tabularx}
\usepackage{array}
\usepackage[all]{nowidow}
\usepackage{microtype}

\DeclareUnicodeCharacter{2218}{\ensuremath{\circ}}

\begin{document}


\firstsection{Introduction}

\maketitle

Telepresence has, since Minsky’s articulation in the 1980s \cite{minsky1980telepresence}, aimed to make remote users feel present in a distant environment. Among the media that support such presence, 360-degree video is especially attractive because it provides continuous panoramic context and can enhance immersion, and has therefore been adopted in a variety of telepresence systems and applications \cite{zhang2018detection, ha2020telepresence}.

However, current 360-degree telepresence pipelines remain constrained by capture, bandwidth, and display limitations. On the capture side, consumer 360-degree cameras can record up to 8K at 30~FPS \cite{ref-insta360x4, ref-insta360x5}, but their live-streaming modes are much more limited, typically providing substantially lower effective detail (e.g., 1280p on the Insta360 X4 and X5). Professional systems can support higher-quality live streaming \cite{ref-insta360pro2}, but at greater cost and reduced portability due to their considerably larger size and weight. On the display side, modern head-mounted displays offer roughly 2K--4K resolution per eye \cite{ref-metaquest3, ref-pimax}, yet their effective angular resolution still falls well below the threshold for eye-limited visual performance (see Section~\ref{subsec:resolution}).

Prior work has explored partial solutions to this problem. MUCVR \cite{li2025mucvr} reduces latency in mobile VR by splitting rendering into dynamic foreground interaction and static background environment components, rendered at the edge and in the cloud, respectively, and then merged before display. ShowMeAround \cite{nassani2021showmearound} addresses the limited detail of live 360$^\circ$ video in virtual tours by allowing presenters to supplement the live stream with pre-recorded high-resolution 360$^\circ$ video and still imagery. These systems suggest two useful ideas: separating static and dynamic content, and augmenting live panoramic views with higher-resolution material when needed.

Motivated by these observations, we focus on a common but underexplored setting: static 360-degree telepresence. Many influential telepresence systems, including Google’s Project Starline \cite{lawrence2024project}, Microsoft’s Viewport \cite{zhang2013viewport}, and Room2Room \cite{pejsa2016room2room}, rely on static capture rigs, whereas moving-camera systems such as JackIn Head \cite{kasahara2015jackin} are less common. Static capture is likewise prevalent in public 360-degree datasets. For example, in the 208-video gaze prediction dataset of Xu \textit{et al.}, 145 videos were recorded with static cameras \cite{xu2018gaze}; Leader360V reports that more than 5,000 of its 10,180 videos are static \cite{zhang2025leader360v}; and 129 of the 150 videos in PanoVOS were captured with static cameras \cite{yan2024panovos, zhang2025leader360v}. Here, “static” refers to a camera fixed relative to its mounting surface, with most of the scene remaining unchanged over~time.

We therefore ask: how can static 360-degree telepresence provide ultra-high-resolution (UHR) detail while preserving live scene dynamics and user-driven inspection? To answer this, we propose a multi-layer system that exploits scene stability to separate panoramic fidelity, dynamic updates, and high-detail viewing in regions of interest. Our hardware combines an 8K panoramic camera with a rotatable 4K pan-tilt-zoom (PTZ) camera. On the software side, we design a three-layer representation:

\textbf{Tile-based ultra-high-resolution panoramic background layer (UHR layer).} Since many telepresence scenes remain largely static, we construct a detailed spherical background offline from an 8K panorama and PTZ-captured 4K views. These views are aligned and fused into the panorama, producing an effective resolution of up to 24K in scanned regions. The resulting background is stored as tiles to support efficient local updates.

\textbf{Dynamic update layer.} To preserve live motion, we apply real-time background matting to the incoming 8K 360-degree stream and composite extracted foreground changes, such as people or moving objects, onto the UHR layer, following Lin \textit{et al.}~\cite{lin2021real}. This layer prioritizes responsiveness over maximum resolution.

\textbf{Region-of-interest 4K layer (ROI layer).} For user-selected regions of interest, a live PTZ stream provides 4K detail for visually demanding tasks such as reading text or inspecting small objects. This layer complements the UHR background and dynamic update with high-detail local viewing.

We evaluate our approach against state-of-the-art super-resolution methods and show that it reconstructs fine details more clearly while preserving immersion. Although the UHR layer is generated offline, the overall design balances panoramic fidelity, live scene updates, and detailed ROI inspection.

The contributions of this paper are as follows:
\begin{itemize}
  \item A three-layer telepresence representation consisting of an offline tile-based UHR background layer, a real-time dynamic update layer, and a live ROI 4K layer.
  \item A practical capture and rendering pipeline that realizes this representation by combining an 8K panoramic camera with a 4K PTZ camera to generate an effective 24K background in scanned regions.
  \item A comparative evaluation against contemporary super-resolution methods and a user study showing improved fine-detail perception and user experience in VR telepresence.
\end{itemize}

\noindent
In summary, the proposed system offers a new approach to reconciling panoramic fidelity with interactivity in static telepresence, paving the way for UHR remote presence under consumer-friendly capture conditions.

\begin{figure*}[h]
    \centering
    \includegraphics[width=\textwidth]{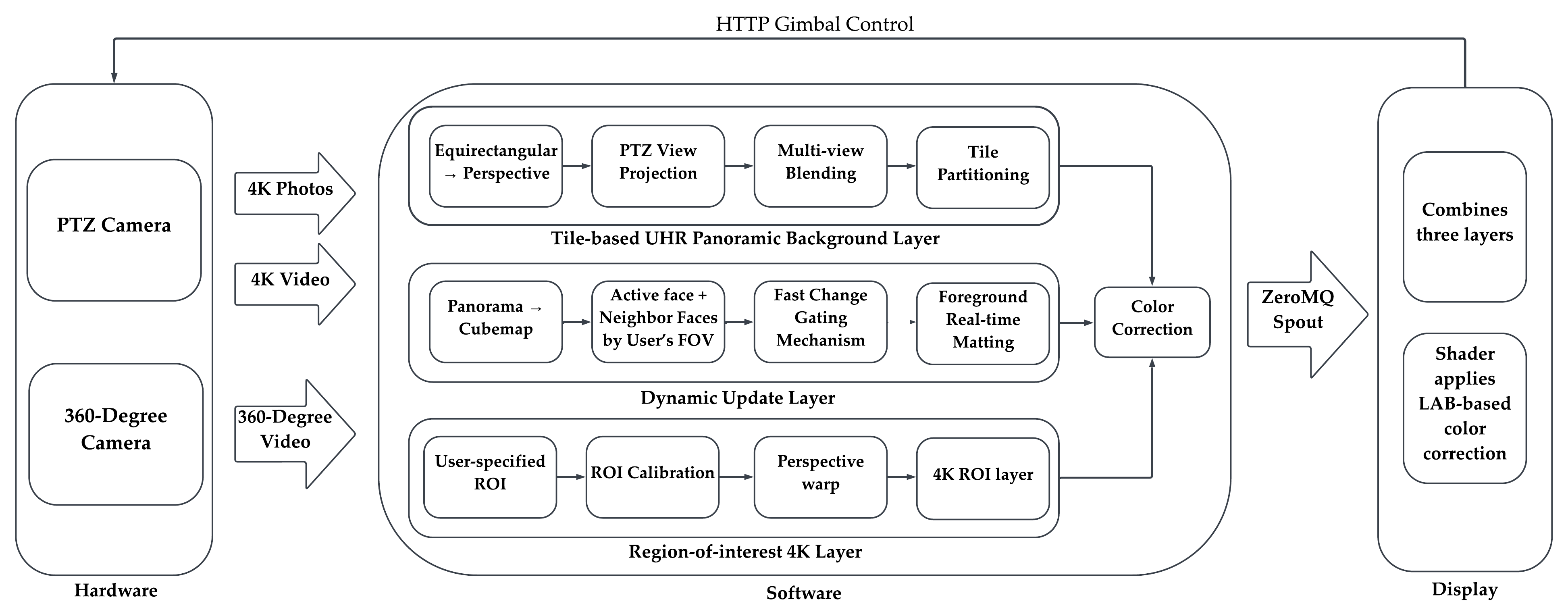}
    \vspace{-15pt}
    \caption{System pipeline diagram.}
    \label{fig:pipeline}
\end{figure*}

\section{Related Work}

In this section we give an overview of related work.

\subsection{Layered and Hybrid 360-Degree Telepresence}

Several systems suggest decomposing immersive telepresence content according to function or quality requirements. MUCVR \cite{li2025mucvr} separates mobile VR rendering into dynamic foreground interaction and static background environment components, with the former rendered at the edge and the latter in the cloud before being merged for display. This design reduces latency and lowers the rendering burden on the client, but it targets interactive mobile VR rather than high-fidelity real-world 360-degree telepresence.

ShowMeAround \cite{nassani2021showmearound} augments live 360-degree video tours with pre-recorded high-resolution 360$^\circ$ videos and still images, allowing viewers to inspect points of interest more clearly than is possible from the live stream alone. While this improves visual detail in guided tours, it does not provide a layered reconstruction pipeline for maintaining a high-resolution panoramic representation over time.

Our system combines insights from both directions. Similar to MUCVR, we separate static and dynamic content into different layers; similar to ShowMeAround, we use a high-resolution auxiliary stream to improve detail beyond the native live 360$^\circ$ video. Unlike either approach, however, we target static-scene VR telepresence with a tile-based UHR panoramic background that can be locally refreshed using ROI-guided PTZ updates.

\subsection{Resolution Requirements}\label{subsec:resolution}
Although 360-degree video can improve situational awareness in telepresence by providing a continuous panoramic view~\cite{narciso2019immersive}, it also imposes stringent resolution requirements. Prior work has estimated these requirements by matching the angular resolution of reference desktop displays in immersive viewing configurations~\cite{shen2014viewing,chi2025modular}. Here we adopt a more direct criterion derived from published visual performance data. Lloyd \textit{et al.} synthesized a series of display evaluations and reported that, with sufficient antialiasing, a pixel pitch of 0.5 to 0.93 arcmin per pixel yields 90\% of peak visual performance for observers with 20/20 or better acuity across a variety of visual tasks, and they recommend this asymptotic performance level as the basis for eye-limited resolution in display design~\cite{lloyd2015practical}. The coarser bound of 0.93 arcmin per pixel corresponds to approximately 64.5 pixels per degree (PPD). More recent psychophysical measurements using purpose-built sliding-resolution hardware indicate that the foveal resolution limit can reach 94 PPD~\cite{ashraf2024resolution}, meaning that performance continues to improve marginally beyond the 64.5 PPD threshold, but the Lloyd \textit{et al.} data establish that the large majority of the perceptual benefit is already realized at that point.

Against this criterion, native 8K 360-degree capture (7680 horizontal pixels over 360 degrees, or 21.3 PPD) falls well short of eye-limited resolution. Our UHR layer reaches an effective horizontal resolution of approximately 24K over the full panorama, corresponding to about 66.7 PPD, which places it just above the 64.5 PPD threshold at which observers achieve 90\% of peak visual performance. These values remain approximate, since perceived quality in HMDs is also influenced by device optics, rendering geometry, the screen-door effect, and vergence--accommodation conflict~\cite{nguyen2020screen,kramida2015resolving,bharadwaj2009accommodative}. Nevertheless, this analysis indicates that the proposed system operates near the empirically grounded target for high-detail VR telepresence, whereas native 8K capture does not.

\subsection{Real-Time High-Resolution Background Matting}
\label{sub:matting}

The dynamic update layer extracts moving objects and people from the 8K 360-degree video stream. This is challenging because most vision models are designed for perspective images, whereas equirectangular projection (ERP) introduces severe geometric distortion and can lead to seams and inconsistent results when processed directly \cite{zhong2025omnisam, shen2022panoformer,zhang2022fast,hilliard2024360u}. We therefore avoid applying matting directly to the full ERP image and instead operate on perspective cube faces.

For foreground extraction, we adopt BackgroundMattingV2 (BMV2) \cite{lin2021real}. BMV2 is a real-time high-resolution matting method that combines a coarse base network with a refinement stage for error-prone regions, enabling a practical balance between quality and efficiency. Compared with earlier background matting approaches such as BGM \cite{sengupta2020background}, it achieves substantially better detail preservation while remaining fully automatic and real time.

Although background-based matting is sometimes considered limited because it requires a reference background image \cite{li2023deep}, this assumption matches our target scenario well: the telepresence scene is largely static, so a clean background reference can be obtained naturally without additional user effort. This makes BMV2 particularly suitable for our system. Other real-time matting methods, such as MODNet \cite{ke2022modnet}, are also possible, but BMV2 integrates more conveniently with our layered architecture because it directly leverages the static-scene prior already assumed by the UHR background.

\section{Methodology}
\label{sec:methodology}

Figure~\ref{fig:pipeline} provides an overview of the complete system workflow, including hardware capture, software processing, and the three output layers.

\subsection{Hardware Design}

\begin{figure}[h]
    \centering
    \includegraphics[width=\columnwidth]{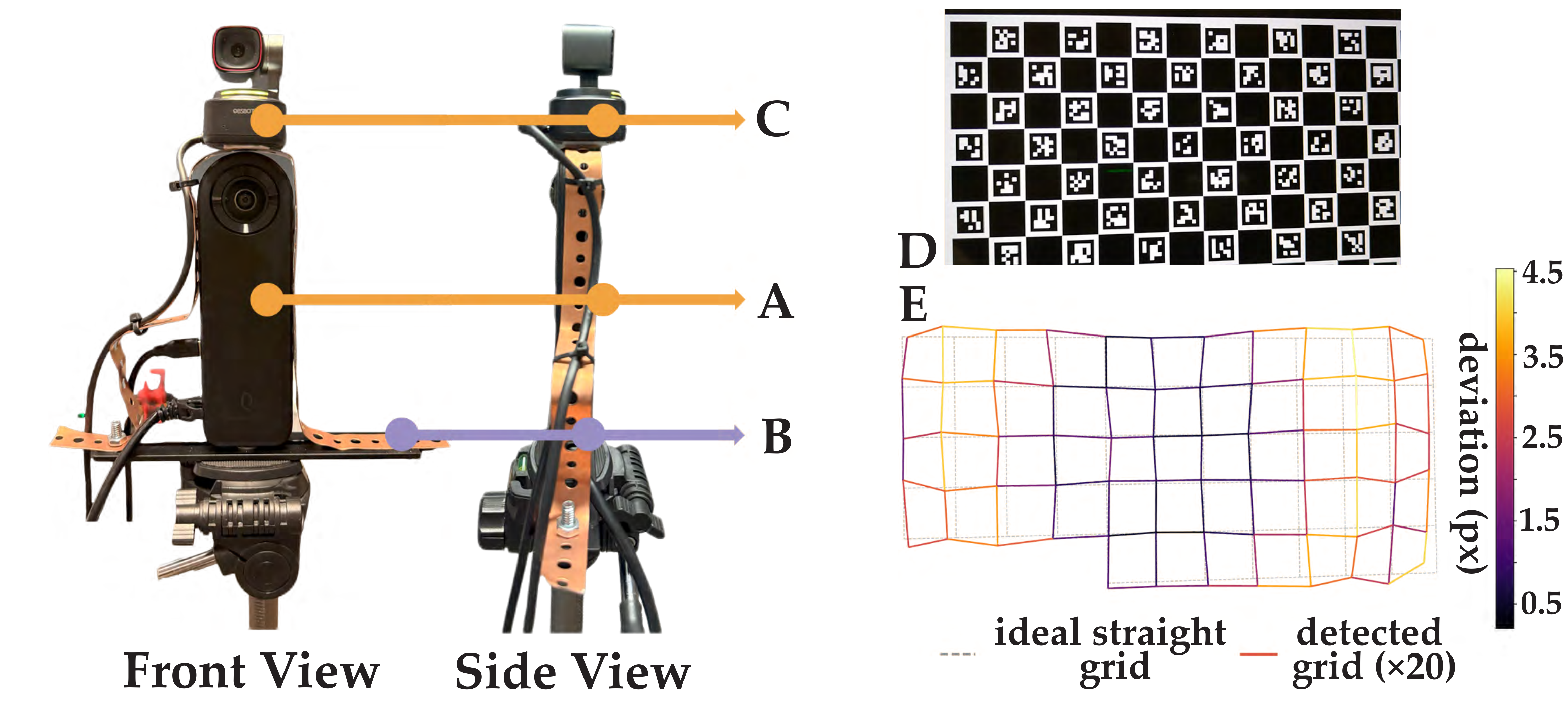}
    \vspace{-15pt}
    \caption{Hardware design of the proposed system. (A) Kandao QooCam 8K Enterprise 360-degree camera, (B) copper strap support structure, (C) OBSBOT Tiny 2 PTZ 4K webcam. (D) A ChArUco board captured by the PTZ camera at its native field of view. (E) Corner deviations from an ideal perspective projection (fitted from central corners only), shown exaggerated for visibility and color-coded by magnitude: the mean deviation grows from 0.88px near the image center to 3.30px toward the corners.}
    \label{fig:hardware}
\end{figure}

The hardware setup consists of three main components, shown in Figure~\ref{fig:hardware}: (A) a Kandao QooCam 8K Enterprise 360-degree camera, (B) a custom copper strap support, and (C) an OBSBOT Tiny 2 PTZ 4K webcam.

The QooCam serves as the panoramic capture unit and supports live streaming at up to 8K and 30~fps via Ethernet. It was selected for its compact form factor, which reduces interference when combined with additional hardware. Mounted above it is the OBSBOT Tiny 2 PTZ webcam, chosen for its small size and controllable directional capture. 
Although its native field of view is 85.5$^\circ$, we found that this setting introduces slight edge distortion. A ChArUco-based measurement (Figure~\ref{fig:hardware} D, E) shows that the deviation from an ideal perspective projection grows from below 1\,px near the image center to over 3\,px toward the corners; such residuals cannot be absorbed by homography-based alignment and accumulate as more PTZ views are stitched. Therefore, to improve geometric consistency when stitching many PTZ views into the UHR layer, we reduce the effective field of view to about 60$^\circ$.
Since a 4K PTZ image then covers about one-sixth of the full 360$^\circ$ panorama, the resulting stitched background corresponds to an effective horizontal resolution of roughly 24K.

The PTZ camera is mounted using a custom copper strap that fits closely around the QooCam while minimizing image occlusion. Both cameras connect to a local computer via wired links (Ethernet for the 360-degree camera and USB for the PTZ camera), which relays the video streams to a workstation for processing and PTZ control. We also tested inverted mounting designs, but the increased distance between the two camera centers reduced feature-matching accuracy. We therefore adopted the strap-based design.

Although our prototype uses a QooCam 8K Enterprise and an OBSBOT Tiny 2, the system is not restricted to these devices. Other 360-degree and PTZ camera combinations could also be used, provided that they offer sufficient resolution and stable integration into the layered telepresence pipeline.

\subsection{Software Design}\label{software}

\subsubsection{Tile-based UHR panoramic background layer}\label{sec:uhr_tile}

\begin{figure}[h]
    \centering
    \includegraphics[width=\columnwidth]{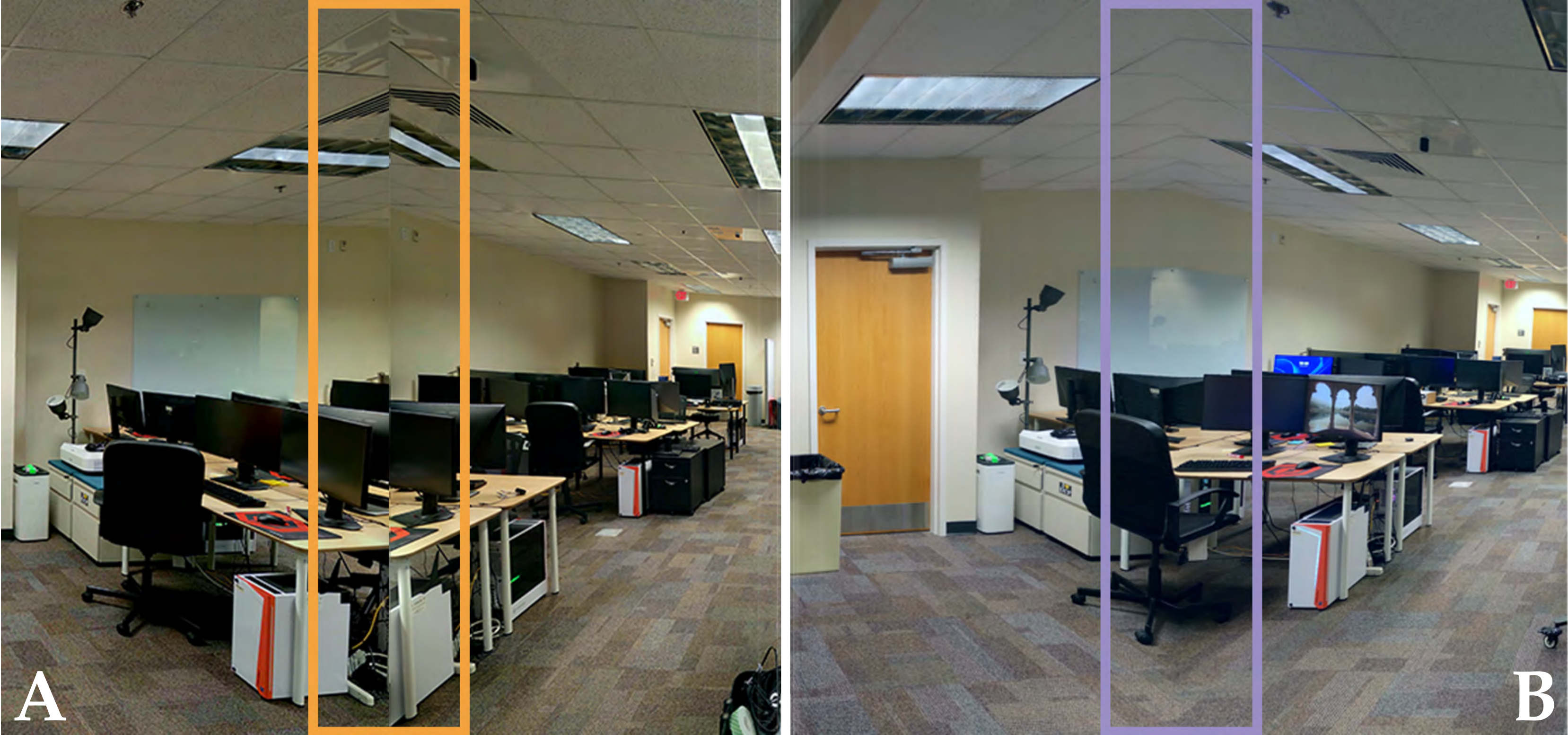}
    \vspace{-15pt}
    \caption{(A) Cubemap matching accumulates cross-face errors, causing visible misalignment. (B) Our virtual-PTZ method enables locally consistent fusion and globally aligned UHR panoramas.}
    \label{fig:panobg}
    \vspace{-4pt}
\end{figure}

Direct feature matching between equirectangular panoramas and high-resolution PTZ captures is unreliable because equirectangular projection introduces strong geometric distortion, especially away from the equatorial region. A straightforward alternative is to convert the panorama into a cubemap and align each face with PTZ captures. However, this breaks global consistency: because the panoramic and PTZ images are acquired with different optics and slightly offset viewpoints, stitching the aligned faces back together can introduce parallax errors and visible seams, as illustrated in Figure~\ref{fig:panobg}.

We therefore use a virtual PTZ that mirrors the pose of the physical PTZ camera. For each PTZ capture, the virtual PTZ renders a perspective view from the equirectangular canvas at the same yaw and pitch, providing a geometrically compatible reference for feature matching. The PTZ image is aligned to this reference, fused by Poisson seamless cloning\cite{perez2003poisson}, and projected back onto the equirectangular canvas. Repeating this process across sampled directions progressively refines the panorama into an ultra-high-resolution background. We considered SIFT\cite{lowe2004distinctive}, LoFTR\cite{sun2021loftr}, and LightGlue+DISK\cite{lindenberger2023lightglue,tyszkiewicz2020disk} for feature matching, and ultimately adopted SIFT for both UHR construction and ROI calibration because robustness was more important than runtime speed in our setting\cite{luo2024comparative}.

The PTZ camera follows a snake-like scan pattern over the usable viewing range. In practice, selecting an appropriate virtual field of view is also important, since autofocus in many webcams can slightly change the effective FOV of the captured images; multiple trials within a reasonable range may therefore be needed to obtain a stable virtual FOV for reliable fusion. The fused panorama is stored as a tile-based UHR background. After fusion, the canvas is resized to the target 24K resolution and partitioned into $24 \times 12$ tiles, each of size $1024 \times 1024$ pixels, allowing later updates to modify only tiles near the region of interest rather than the full panorama. In our implementation, each UHR panorama is generated from approximately 50 PTZ photos in about 25 minutes.

\subsubsection{Dynamic Update Layer}
\label{sub:du}

The purpose of this layer is to extract and stream moving people or objects from the panoramic video in real time. Operating directly on full 8K equirectangular frames is impractical: the pixel count is too high for efficient processing, and, as discussed in Section~\ref{sub:matting}, the distortions of equirectangular projection reduce the performance of matting models that are typically designed for perspective images.

To address this, we first convert the panoramic stream into a cubemap. At runtime, depending on the user’s viewing direction, the system processes the corresponding cube face, its two adjacent side faces, and the bottom face. This design improves visual continuity, since most VR headsets have a total FOV larger than $90^\circ$, and relevant motion may extend across face boundaries or into the ground-facing region. The top face is excluded because, in our setup, it mainly captures the PTZ camera base.

Before performing expensive matting, we apply a fast change-gating mechanism. Each selected face is downsampled, converted to grayscale, and compared with an exponentially averaged reference frame. If only minor changes are detected, the face is skipped; otherwise, it proceeds to high-resolution processing. This step avoids unnecessary computation while preserving responsiveness to meaningful foreground motion.

Faces that pass the gate are processed with a real-time background matting model following Lin \textit{et al.}\ \cite{lin2021real}. The network predicts an alpha matte and foreground colors from the cube face and its static background reference. The extracted foreground is then blended back using Poisson seamless cloning to improve visual smoothness. For directions without detected motion, transparent placeholders are sent so that the renderer always receives a complete set of inputs.

At runtime, the equirectangular video is decoded on the GPU, converted into cube faces, and evaluated by the gating step. Only faces with significant changes are forwarded to the matting model, converted to RGBA outputs, and streamed to the renderer, where they are composited over the UHR background. This design enables efficient foreground extraction without processing the full 8K panorama while maintaining visual continuity and real-time responsiveness.

\subsubsection{Region-of-interest 4K layer}

This layer allows users to select regions of interest for high-detail inspection. ROI selection is by design decoupled from any specific input modality: the system exposes a programmatic interface that any external input source, such as controller pointing, gaze, or a desktop client, can drive; in our user study (Section~\ref{sec:experiment}), the ROI was fixed to a predefined region for experimental control. Once selected, the PTZ camera is directed toward the target region, and a one-time SIFT-based homography calibration is performed against a GPU-projected perspective view of the panorama using CuPy~\cite{cupy_learningsys2017}. Subsequent frames are aligned using the estimated homography without repeated feature matching. 
The ROI layer is not coupled in real time with HMD movements; instead, each transition is treated as a discrete event: when the user commits a new target region, the PTZ camera is mechanically reoriented and the alignment is re-estimated, during which the ROI view is hidden and the underlying UHR background remains visible, and the live stream reappears only once the pipeline has stabilized.

\begin{figure}[t]
    \centering
    \includegraphics[width=\columnwidth]{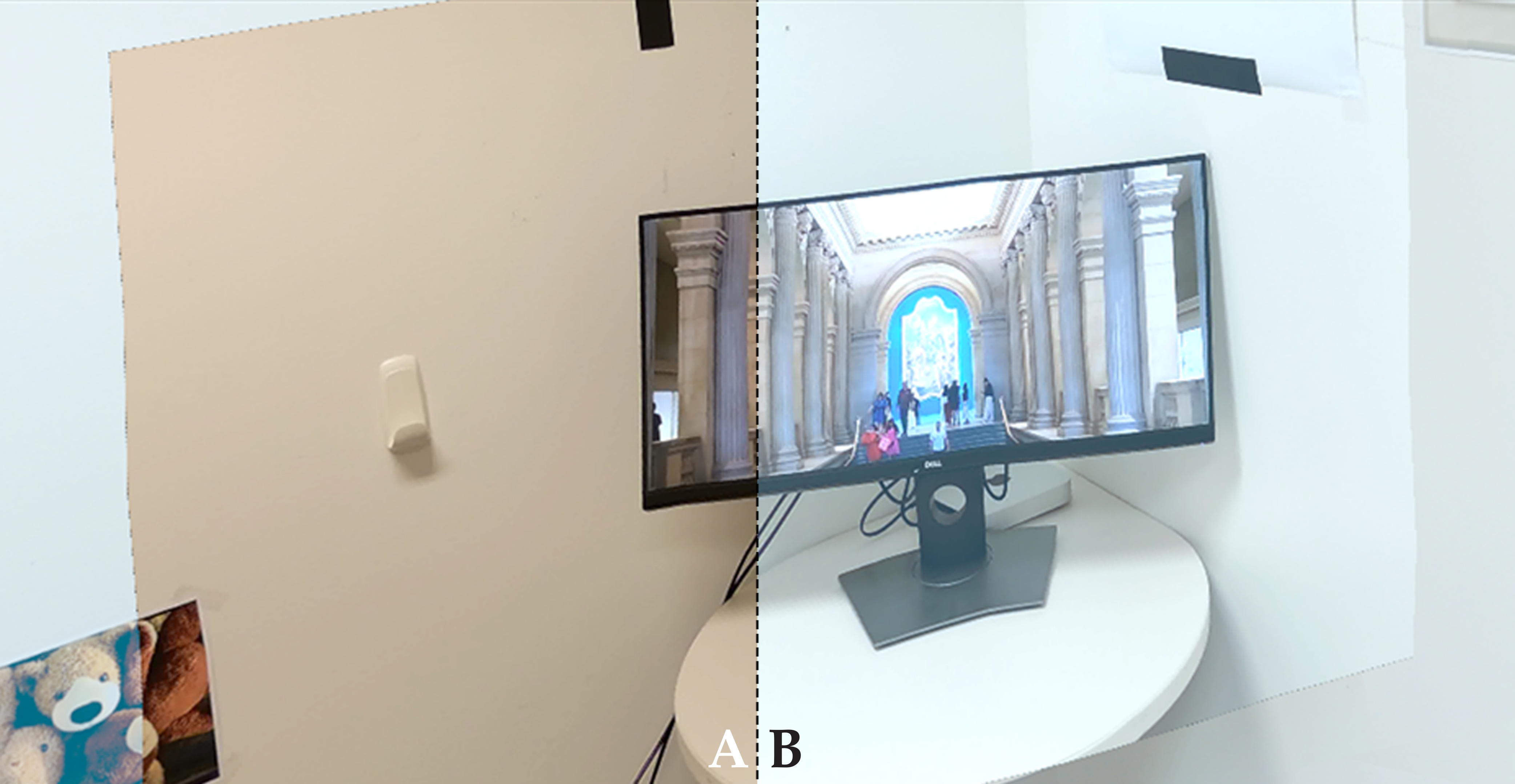}
    \vspace{-15pt}
    \caption{Color correction comparison: (A) original overlay with visible discrepancies; (B) result after color correction.}
    \label{fig:color_transfer}
\end{figure}

The ROI layer provides a live 4K PTZ stream, transmitted directly to Unity via Spout\cite{ref-spout}, while ZeroMQ\cite{ref-zeromq} is used only for lightweight placement metadata and control messages. This avoids large per-frame transfers through the messaging channel and enables efficient real-time display of the ROI content. To reduce visible color discrepancies between the two cameras, we apply an offline color transfer method based on Reinhard \textit{et al.}~\cite{reinhard2002color}, with precomputed correction parameters applied in Unity shaders; this is used for both the ROI and dynamic update layers (Figure~\ref{fig:color_transfer}). The same ROI input is also used to refresh the tiled UHR background locally, following the same matching and fusion procedure used during tile generation, but restricted to the tiles intersecting the ROI and their immediately surrounding ring rather than the full panorama.

Notably, this image-based alignment, together with the offline tile registration in Section~\ref{sec:uhr_tile}, is all the calibration the system requires: both stages align PTZ content against the shared panoramic reference, so no explicit extrinsic calibration between the two cameras is needed.

Overall, the ROI layer supports both live 4K inspection and localized refinement of the UHR background.

\section{System Evaluations}

\subsection{Evaluation Setup}
\label{sub:eva_set}

\begin{figure*}[tb]
  \centering
  \includegraphics[width=\textwidth]{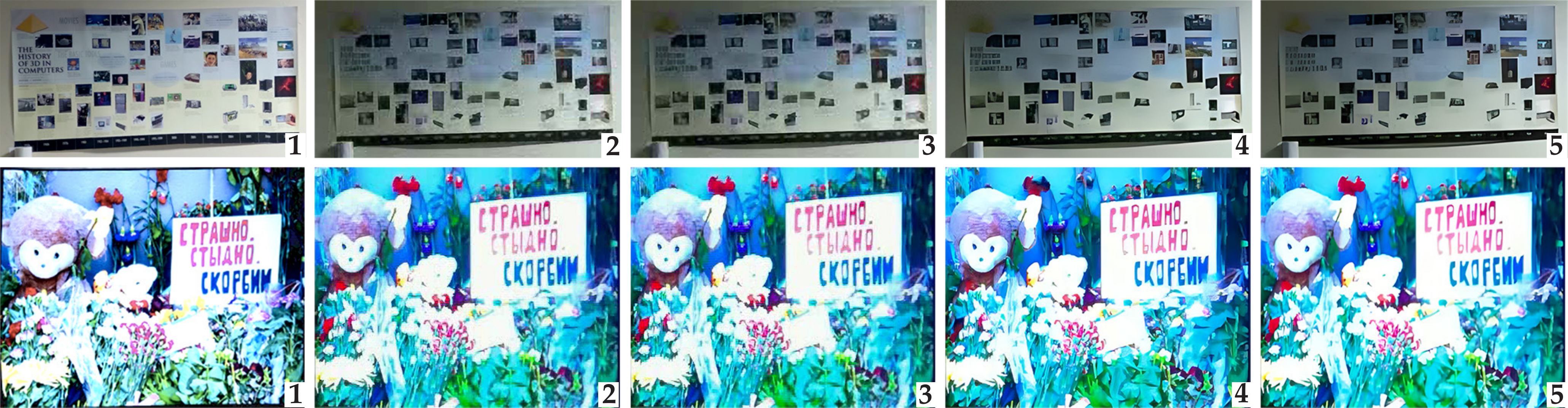}
  \vspace{-15pt}
  \caption{Visual comparison arranged in two groups: the upper group corresponds to the UHR layer, and the lower group corresponds to the ROI layer. Within each group, five columns are shown: (1) our method, (2) DRCT, (3) HAT, (4) RealBasicVSR, and (5) Real-ESRGAN}
  \label{fig:sr_comparison}
\end{figure*}

Our evaluation setup consists of two cameras, a relay computer, and a high-performance workstation. Panoramic capture is provided by a Kandao QooCam 8K Enterprise at $7680 \times 3840$ and 30~fps using H.265 encoding at 30~Mb/s. Directional capture is provided by an OBSBOT Tiny~2 PTZ 4K webcam, streaming at $3840 \times 2160$ and 30~fps.

Rather than connecting the cameras directly to the workstation, both streams are first routed through a relay computer, which forwards them over the local network via ZeroMQ and also receives PTZ control commands. This design better reflects remote telepresence deployment. The relay computer is equipped with an AMD Ryzen~9~5900HS CPU, 32~GB RAM, and an NVIDIA RTX~3080 GPU.

All processing described in Section~\ref{software} is performed on a workstation with an Intel Core i7-14700KF CPU, 64~GB RAM, and an NVIDIA RTX~4090 GPU. Rendering is implemented in Unity3D (version~2023.2.20f1), and all super-resolution baselines are also executed on the same workstation.

\subsection{Evaluation Scenarios and Metrics}
We evaluate the system in a $25$\,m\,$\times$\,$10$\,m indoor space with ceiling lights as illumination and occasional point sources such as monitors.

\begin{table}[tb]
  \caption{Dynamic update layer per-frame runtime.}
  \label{tab:layer2}
  \scriptsize
  \centering
  \begin{tabu} to \linewidth {X[c] X[c]}
  \toprule
   Stage & Time (ms) \\
  \midrule
   Frame upload to GPU & 7.66 \\
   Gating & 5.43 \\
   Equirect $\rightarrow$ Cubemap resampling & \textbf{27.50} \\
   Matting inference & \textbf{25.35} \\
   Output & 8.54 \\
  \midrule
   \textbf{Total time} & 74.48 \\
  \midrule
   Effective FPS & 13.43 \\
   Actual FPS & \textbf{12.76} \\
  \bottomrule
  \end{tabu}
\end{table}

\subsubsection{Performance}

Table~\ref{tab:layer2} reports mean per-frame times in the dynamic update layer. The main runtime cost is shared between resampling the 8K panorama into cubemap faces and the matting stage, which take comparable time. Together, they dominate the per-frame processing cost.

For the ROI layer, steady-state playback matched the 30\,FPS source stream. The main per-update cost comes from computing the projected ROI region in panorama space. Latency was measured visually in an end-to-end manner. Because the pipeline spans multiple systems, direct internal timing was difficult to obtain. The operator activated a small light in the scene while noting the trigger time on a synchronized millisecond clock, and the receiving display was screen-recorded. Latency was then estimated by comparing the trigger time with the first frame in which the light became visible on the recorded display. Measured in this way, the latency was 1.41\,s for the dynamic update layer and 1.47\,s for the ROI layer.

These frame rates refer only to the individual layers rather than the overall VR application, which remained stable at 90\,FPS throughout runtime. As also reflected in Section~\ref{sub:addition}, this was associated with only a slight increase in sickness symptoms for participants. Since SIFT-based calibration is performed only when the user repositions the ROI window, it does not affect steady-state playback.

\subsubsection{Comparison with Super-Resolution Methods}
\label{sub:sr}

We compared our system with four representative super-resolution methods---DRCT~\cite{hsu2024drct}, HAT~\cite{chen2023hat}, RealBasicVSR~\cite{chan2022investigating}, and Real-ESRGAN~\cite{wang2021real}---using their official pre-trained $\times$4 models. The aim was to contrast learned upscaling with our capture-based multi-layer design.

As noted in Section~\ref{sub:du}, equirectangular projection introduces distortions that reduce the performance of models designed for perspective images. We therefore first converted each 8K panorama into a cubemap before applying the SR methods. Figure~\ref{fig:sr_comparison} shows comparisons for the UHR and ROI layers; the dynamic update layer is omitted because it does not perform resolution enhancement.

In both groups, our system preserves fine details more faithfully. In the UHR layer, small distant text remains clearer, whereas SR outputs often smooth details or hallucinate content. In the ROI layer, our system has a direct advantage because it streams real 4K camera content rather than synthesizing it. Overall, SR can improve apparent sharpness, but our capture-based design provides more reliable scene detail for telepresence.

\section{Experiment}
\label{sec:experiment}

In this section, we describe the experiment we conducted to evaluate our system.
The experiment was approved by the Institutional Review Board (IRB) of our university.

\subsection{Participants}
\label{sec:participants}

We recruited 21 participants for our experiment (15 male and 6 female; ages 20 to 46, $M$\,=\,27.3, $SD$\,=\,6.7).
The participants were members of the local university community.
All of the participants had normal or corrected-to-normal vision.
We verified 20/20 vision for all participants before the experiment using a standardized Snellen chart~\cite{elliott2020clinical}.
We asked our participants to indicate their prior experience with the technologies we used in this study. 1 participant had little or no prior experience with VR devices, 10 reported moderate experience, and 10 indicated substantial experience. With respect to telepresence systems, 7 participants reported minimal prior exposure, 9 had moderate familiarity, and 5 described themselves as having considerable experience. For 360-degree video viewing, 6 participants indicated limited familiarity, 9 reported a moderate level of experience, and 6 reported extensive experience.

\subsection{Material}
\label{sec:material}

\begin{figure}[t]
\centering
\includegraphics[width=\columnwidth]{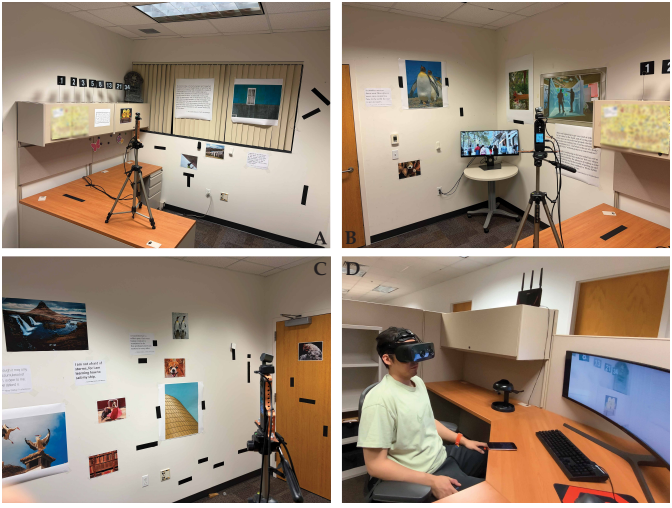}
\vspace{-15pt}
\caption{Experimental setup. (A, B, C) Views of the remote room used in the study. (D) Illustration of the viewing setup. \textit{Where's Waldo?} is blurred for copyright reasons.}
\label{fig:setup}
\vspace{-10pt}
\end{figure}

\begin{figure*}[ht]
\centering
\subfigure[]{\includegraphics[width=0.3\textwidth]{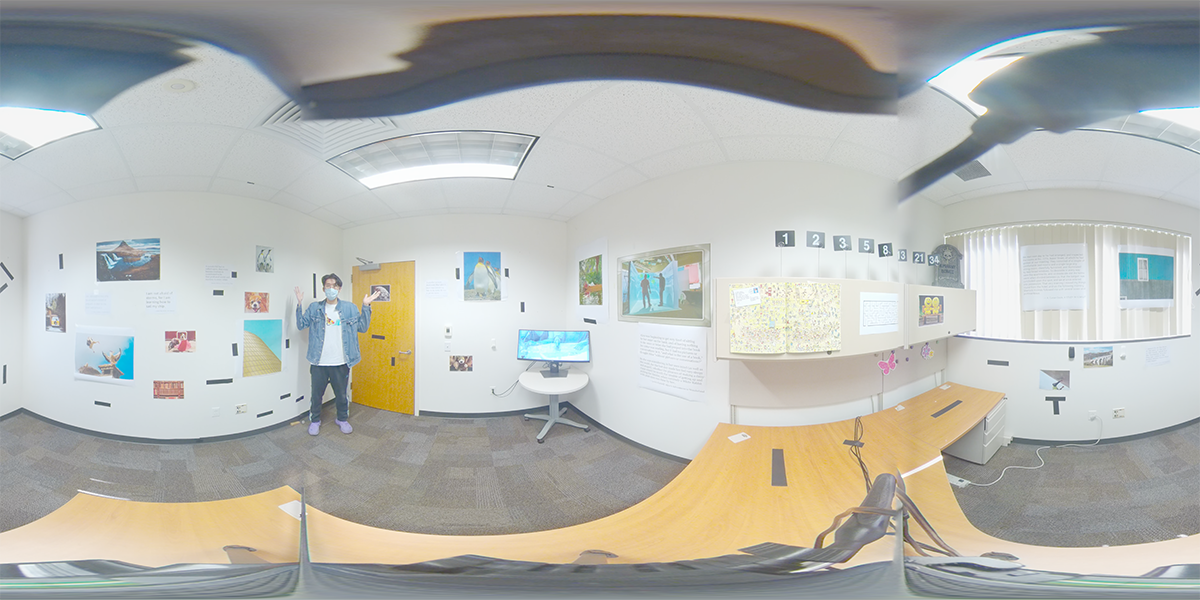}}\quad
\subfigure[]{\includegraphics[width=0.3\textwidth]{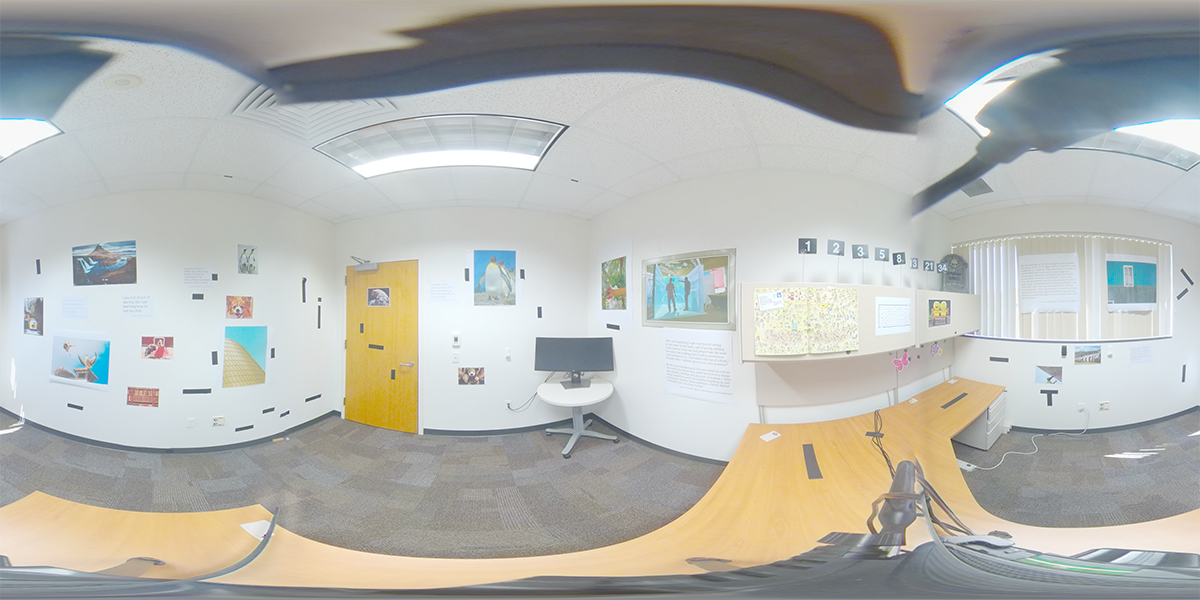}}\quad
\subfigure[]{\includegraphics[width=0.3\textwidth]{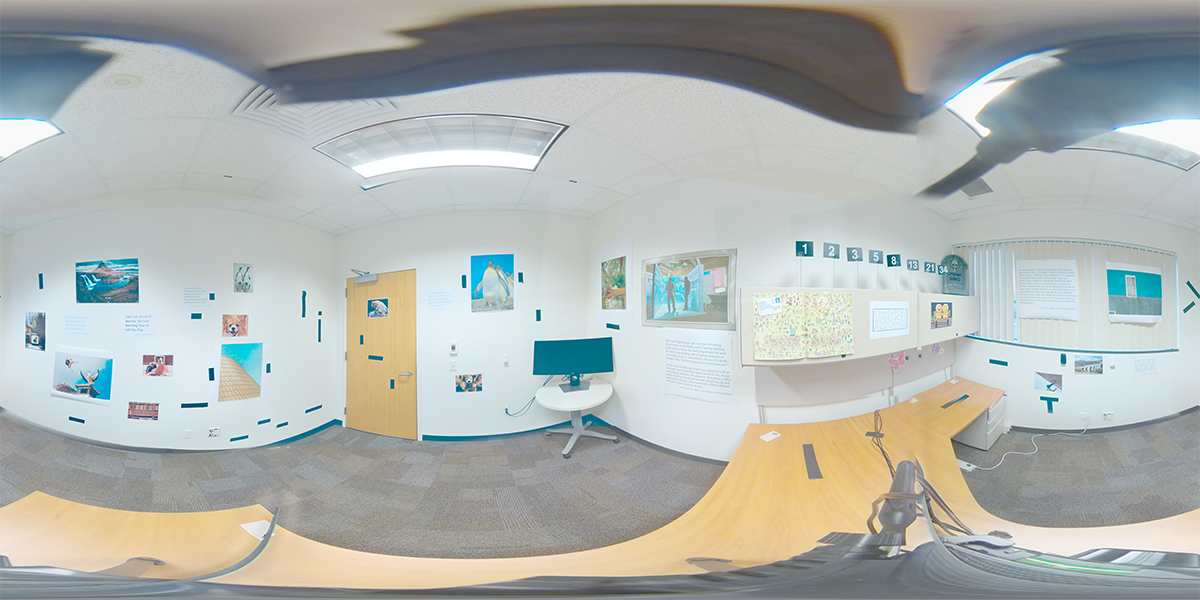}}
    \subfigure[]{\includegraphics[width=0.3\textwidth]{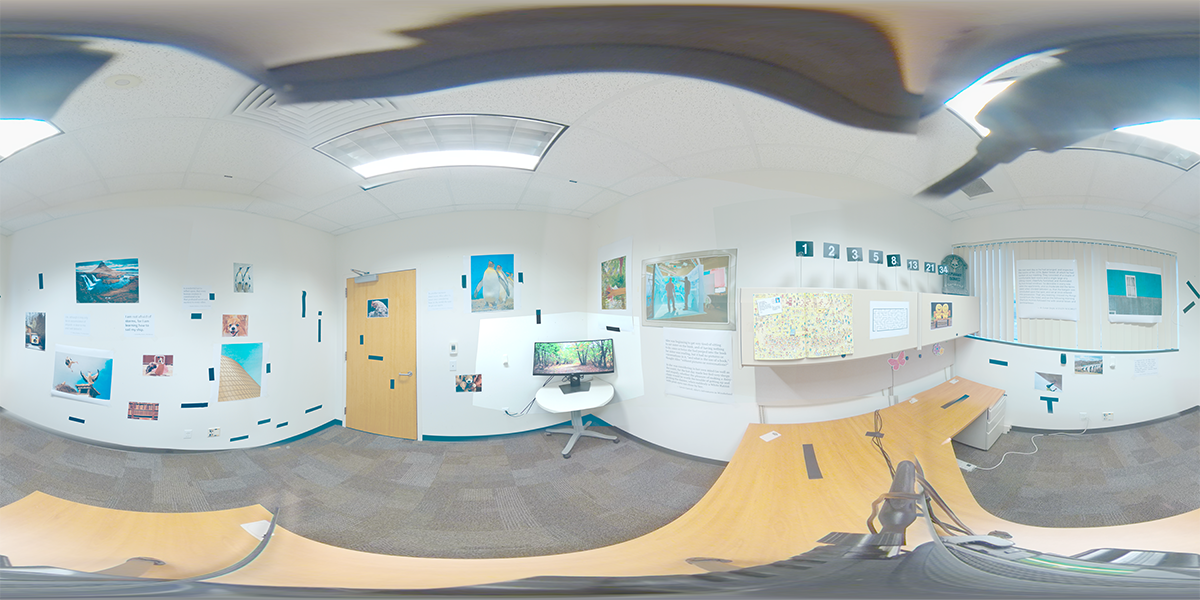}}\quad
\subfigure[]{\includegraphics[width=0.3\textwidth]{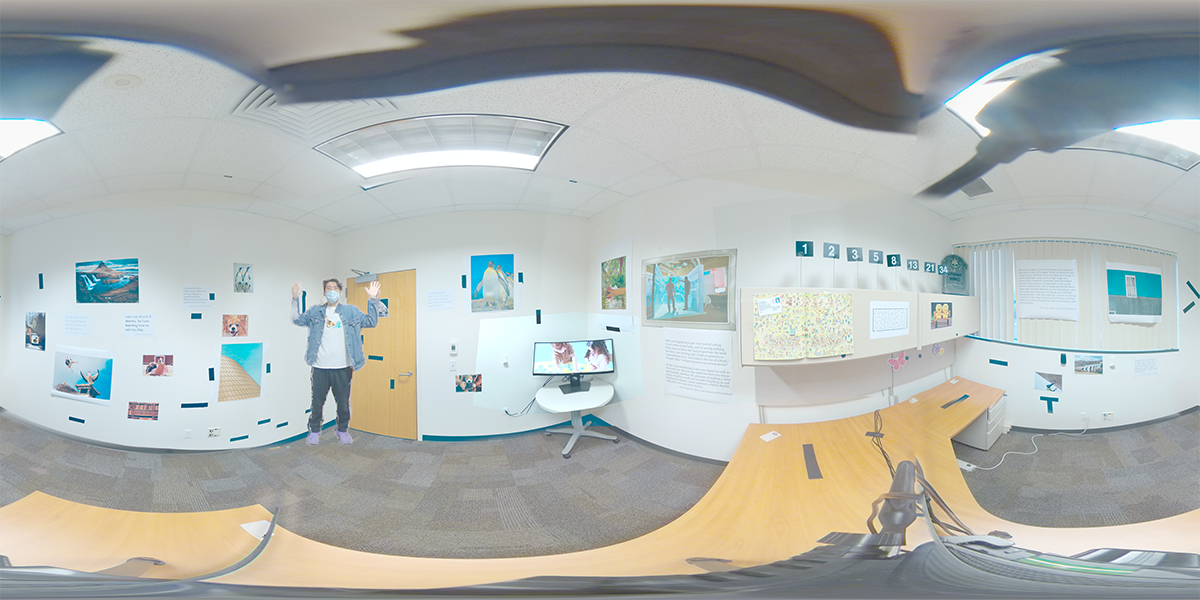}}
\vspace{-6pt}
\caption{Visual stimuli used in the five conditions in the experiment: (a) 8DMR, (b) 8SNN, (c) 24SNN, (d) 24SNR, and (e) 24SMR.}
\label{fig:conditions}
\end{figure*}

Our experiment setup employed the system described in Section~\ref{sec:methodology}. It was installed in a separate room from the participants (see Figure~\ref{fig:setup}D). Participants were seated and asked to wear a Varjo XR-3 HMD. The HMD provides a resolution of 1920 $\times$ 1920 pixels per eye within a 27 $\times$ 27 degree focus area (70 PPD), and 2880 $\times$ 2720 pixels per eye over a 115 $\times$ 90 degree field of view at 90 Hz~\cite{ref-varjo, ref-vrcompare-varjo}. Through the HMD, participants viewed a live stream of the remote room (see Figure~\ref{fig:setup}(A, B, C) and Figure~\ref{fig:conditions}). The other hardware components were identical to those described in Section~\ref{sub:eva_set}.

The remote room was an ordinary office-sized space of approximately 3\,m\,$\times$\,3\,m. To create visually diverse content, we attached printed images from the DIV2K dataset~\cite{agustsson2017ntire} as well as other copyright-free images to the walls. We also displayed excerpts from public-domain books obtained from Project Gutenberg~\cite{ref-gutenberg}. Near the camera, we placed an open \textit{Where's Waldo?} book and a printed maze image, together with several number cards positioned nearby. In one corner of the room, a monitor continuously played copyright-free video clips containing a range of content types, including natural scenes, outdoor environments, and prominent textual elements. To make rendering artifacts such as smearing, and particularly object misalignment, more visually salient, we additionally placed strips of black tape on otherwise blank wall areas.

Before each session, our system scanned the entire room and generated the tile-based UHR panoramic background layer. The PTZ camera was aimed at the monitor region throughout the experiment, continuously transmitting the corresponding ROI video stream. No manual interaction or viewpoint control was required from participants.

\subsection{Methods}
\label{sec:methods}

We used a within subjects design for this experiment and compared five conditions (see Table~\ref{tab:selection} and Figure~\ref{fig:conditions}). The conditions were designed to differ in specific visual components, allowing us to isolate the contribution of different scene elements across representations. In particular, the set of conditions varied in whether they provided a dynamic or static background, whether a moving subject was present, and whether an ROI region was shown.
Among them, \textbf{8DMR} consisted of an \underline{8}\,K \underline{D}ynamic background with a \underline{M}oving subject and a visible \underline{R}OI region, and thus was essentially a native 8\,K video stream. \textbf{24SMR}, in contrast, consisted of a \underline{24}\,K \underline{S}tatic background (UHR layer) with a \underline{M}oving subject (dynamic update layer) and a visible \underline{R}OI region (ROI layer), corresponding to our proposed system.

This condition set spans the endpoints that are currently realizable: 8DMR represents the practical ceiling of native 360-degree live streaming, while 24SMR represents our full system. A 24K dynamic baseline, which would fully disentangle resolution, layering, and background dynamics, is not obtainable, as to our knowledge no commercial 360-degree camera produces a live stream at such a resolution---precisely the gap our system addresses. The intermediate conditions vary one component at a time (8SNN vs.\ 24SNN isolates background resolution, 24SNN vs.\ 24SNR the ROI layer, and 24SNR vs.\ 24SMR the dynamic update layer); other 8K-static combinations were not tested, as the dynamic update and ROI layers are integrated with the UHR background construction and would correspond to neither our system nor any standard baseline. Consequently, only comparisons between adjacent conditions can be attributed to individual components, whereas the 8DMR--24SMR comparison reflects the combined effect of the full system. All conditions in this experiment were presented to participants in random order.

\begin{table}[h!]
  \caption{Conditions in the experiment. ``$\times$'' indicates that the component is included in the condition, and ``--'' that it is not.}
  \label{tab:selection}
  \scriptsize
  \centering
  \begin{tabularx}{\linewidth}{
>{\centering\arraybackslash}X
>{\centering\arraybackslash}X
>{\centering\arraybackslash}X
>{\centering\arraybackslash}X
>{\centering\arraybackslash}X
}
  \toprule
  Condition
  & \multicolumn{1}{c}{\makecell{Background\\Resolution}}
  & \multicolumn{1}{c}{\makecell{Background\\Type}}
  & \multicolumn{1}{c}{\makecell{Moving\\Subject}}
  & \multicolumn{1}{c}{\makecell{ROI}} \\
  \midrule
\textbf{8DMR} & 8K & Dynamic & $\times$ & $\times$ \\
\textbf{8SNN} & 8K & Static & $-$ & $-$ \\
\textbf{24SNN} & 24K& Static &$-$ & $-$ \\
\textbf{24SNR} &  24K& Static & $-$ & $\times$ \\
\textbf{24SMR} & 24K & Static & $\times$ & $\times$ \\
  \bottomrule
  \end{tabularx}
\end{table}


\subsubsection{Procedure}

Upon arrival, participants first provided verbal consent to participate in the experiment.
Participants then performed a visual acuity test using a Snellen chart to ensure they had 20/20 or better vision.
The experimenter then asked them to fill out a questionnaire capturing demographic information and prior experience. No formal training phase was conducted beyond a brief orientation to the HMD. Participants then completed the five experimental conditions, each lasting approximately 1.5 minutes and followed by a questionnaire. The total time per participant, including questionnaires, was approximately 30 minutes. During each condition, participants remained seated and passively viewed the live-streamed VR representation of the remote office environment through the HMD. No interaction with the system was required; participants were instructed to observe the details of the scene, including pictures and text on the walls and the objects in the room. To introduce live dynamic content, a member of the research team moved within the remote room during the presentation. When enabled, the ROI remained fixed on a predefined area. Condition order was randomized across participants.

At the end of the experiment, participants completed a post-study questionnaire, were debriefed by the experimenter, and received monetary compensation for their participation.

\begin{table}[t!]
\centering
\caption{Single-Item Questionnaires on a 7-point Likert scale (1=strongly disagree, 7=strongly agree) we used in the experiment.}
\vspace{-1pt}
\label{tab:single_item_questions}
\renewcommand{\arraystretch}{1.1}
\resizebox{\columnwidth}{!}{\begin{tabular}{c|l}
\hline
\hline
\textbf{ID} & \textbf{Question}\\ \hline
VD1 & I was able to clearly see fine visual details in the scene. \\ \hline 
VD2 & Text or signs in the scene (if present) were easy to read. \\ \hline 
VD3 & The overall image appeared sharp rather than blurry. \\ \hline 
RE1 & The scene appeared visually realistic and natural. \\ \hline 
RE2 & This viewing helped me understand what was\\
& happening in the scene. \\ \hline 
ART & I noticed no visible artifacts, smearing, or object\\
& misalignment during the viewing. \\ \hline 
ROI & The video displayed on the monitor was easy to see\\
& and understand.\\ \hline 
DU & The moving people or objects appeared visually\\
& consistent with the background.\\
\hline \hline
\end{tabular}}
\vspace{-5pt}
\end{table}

\subsubsection{Measures}

We collected the following subjective measures per condition:

\begin{itemize}
\item \textbf{User Experience Questionnaire Short (UEQ-S)}: We used the short version of the UEQ~\cite{hinderks2017design} to assess hedonic and pragmatic qualities as well as overall user experience in each condition.

\item \textbf{Igroup Presence Questionnaire (IPQ)}: We used the 14-item IPQ~\cite{schubert2001experience} to assess participants' sense of feeling present across conditions.
It measures the four sub-scales general presence, spatial presence, involvement, and experienced realism.

\item \textbf{Single Item Questionnaires}: We included eight further single-item questions on a 7-point Likert scale (1=strongly disagree, 7=strongly agree) as shown in Table~\ref{tab:single_item_questions}.
\end{itemize}

\noindent
Further, we asked participants to fill out general questions at the end of the experiment as well as fill out the simulator sickness questionnaire before and after the experiment:

\begin{itemize}
\item \textbf{General Questions (GQ)}: We asked participants to complete four general single-item questions on a 7-point Likert scale (1=strongly disagree, 7=strongly agree) at the end of the experiment as shown in Table~\ref{tab:general_questions}.

\item \textbf{Simulator Sickness Questionnaire (SSQ)}: Participants completed the SSQ~\cite{kennedy1993simulator} before and after the experiment, which gives an indication of the increase in sickness symptoms during the experiment.
\end{itemize}


\begin{table}[t!]
\centering
\caption{General Questions on a 7-point Likert scale (1=strongly disagree, 7=strongly agree).}
\vspace{-1pt}
\label{tab:general_questions}
\renewcommand{\arraystretch}{1.1}
\resizebox{\columnwidth}{!}{\begin{tabular}{c|l}
\hline
\hline
\textbf{ID} & \textbf{Question}\\ \hline
GQ1 & Improving visual detail made the viewing\\
& experience more useful for understanding the scene.\\ \hline 
GQ2 & I was willing to tolerate minor visual artifacts or\\
& inconsistencies in exchange for clearer visual details.\\ \hline 
GQ3 & In real-world telepresence scenarios (e.g., remote\\
& classes or meetings), being able to clearly see fine \\
& visual details would be important to me.\\ \hline 
GQ4 & The viewing experience demonstrated the potential\\
& value of higher visual detail for telepresence\\
& applications.\\ \hline 
\hline
\end{tabular}}
\vspace{-5pt}
\end{table}

\subsection{Results}
\label{sec:results}

In this section, we present the results of our experiment.
We report the results of a one-way repeated measures ANOVA comparing the five conditions.
We also performed post-hoc pairwise comparisons with Bonferroni correction.
We verified the assumptions of these statistical tests~\cite{verma2019testing}.
This includes the \emph{normality} of the data, which we confirmed with Shapiro-Wilk normality tests, and \emph{sphericity}, which we tested for with Mauchly's test of sphericity.
If Mauchly's test indicated that the assumption of sphericity was not supported, we corrected for it using Greenhouse-Geisser estimates of sphericity.

\subsubsection{User Experience}
\label{sub:ueq}

Figure~\ref{fig:results_standard_questionnaires}(a) shows the results for the UEQ-S.

We found a significant main effect of the conditions on \emph{Hedonic Quality}, $F(2.8,55.5)$\,=\,8.12, \textbf{\itshape{p}\,$<$\,0.001}, $\eta^2_p$\,=\,0.29.
Our post-hoc tests showed that the scores for 8SNN were significantly lower than for 8DMR, 24SNR, and 24SMR (all $p$\,$<$\,0.05).

We also found a significant main effect of the conditions on \emph{Overall Usability}, $F(4,80)$\,=\,5.41, \textbf{\itshape{p}\,$<$\,0.001}, $\eta^2_p$\,=\,0.21.
Our post-hoc tests showed that the scores for 8SNN were significantly lower than for 24SNN, 24SNR, and 24SMR (all $p$\,$<$\,0.05).

We found no significant main effect of the conditions on \emph{Pragmatic Quality}, $F(4,80)$\,=\,1.55, $p$\,=\,0.20, $\eta^2_p$\,=\,0.07.

\subsubsection{Presence}
\label{sub:ipq}

\begin{figure*}[ht]
\centering
\subfigure[]{\includegraphics[height=0.7\columnwidth]{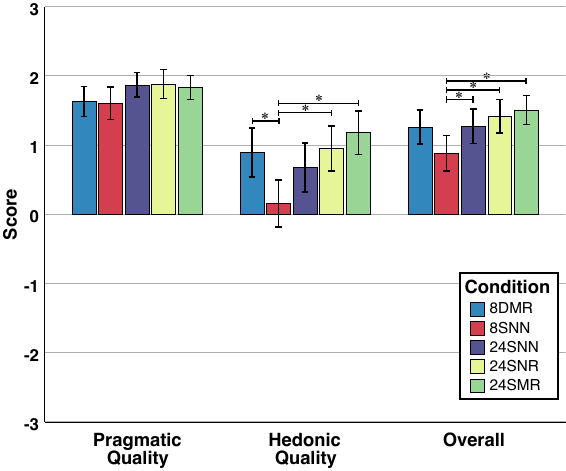}}\quad\quad\quad
\subfigure[]{\includegraphics[height=0.7\columnwidth]{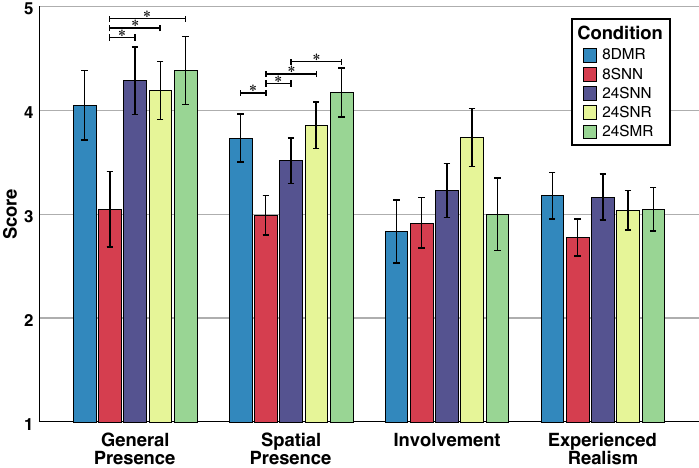}}
\vspace{-8pt}
\caption{Results for the standard questionnaires: (a) User Experience Questionnaire -- Short, and (b) Igroup Presence Questionnaire. Higher is better. The vertical error bars show the standard error. The horizontal bars indicate pairwise significance at the 5\% significance level.}
\label{fig:results_standard_questionnaires}
\end{figure*}

Figure~\ref{fig:results_standard_questionnaires}(b) shows the results for the IPQ.

We found a significant main effect of the conditions on \emph{General Presence}, $F(4,80)$\,=\,6.63, \textbf{\itshape{p}\,$<$\,0.001}, $\eta^2_p$\,=\,0.25.
Our post-hoc tests showed that the scores for 8SNN were significantly lower than for 24SNN, 24SNR, and 24SMR (all $p$\,$<$\,0.05).

We also found a significant main effect of the conditions on \emph{Spatial Presence}, $F(4,80)$\,=\,8.35, \textbf{\itshape{p}\,$<$\,0.001}, $\eta^2_p$\,=\,0.29.
Our post-hoc tests showed that the scores for 8SNN were significantly lower than for 8DMR, 24SNN, and 24SNR, and the scores for 24SNN were significantly lower than for 24SMR (all $p$\,$<$\,0.05).

We further found a significant main effect of the conditions on \emph{Involvement}, $F(4,80)$\,=\,3.13, \textbf{\itshape{p}\,=\,0.019}, $\eta^2_p$\,=\,0.14.
We found no significant post-hoc tests.

We found no significant main effect of the conditions on \emph{Experienced Realism}, $F(4,80)$\,=\,1.42, $p$\,=\,0.24, $\eta^2_p$\,=\,0.07.





\subsubsection{Single-Item Questionnaires}
\label{sub:siq}

\begin{figure*}[ht]
\centering
\includegraphics[height=0.7\columnwidth]{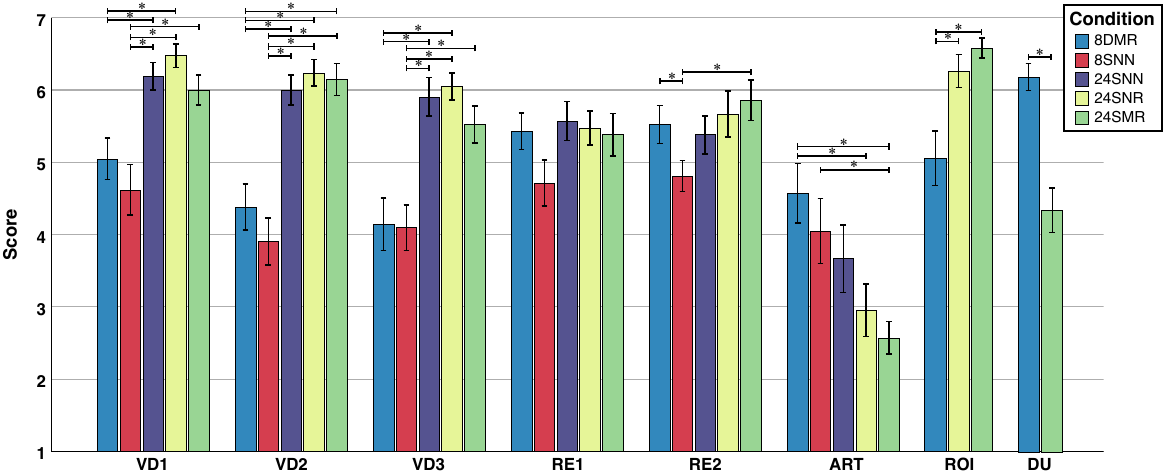}
\vspace{-6pt}
\caption{Results for the single-item questionnaires. Higher is better. The vertical error bars show the standard error. The horizontal bars indicate pairwise significance at the 5\% significance level.}
\label{fig:results_singleitem}
\end{figure*}

Figure~\ref{fig:results_singleitem} shows the results for these scales.

We found a significant main effect of the conditions on \emph{VD1}, $F(4,80)$\,=\,13.20, \textbf{\itshape{p}\,$<$\,0.001}, $\eta^2_p$\,=\,0.40.
Our post-hoc tests showed that the scores for 8DMR were significantly lower than for 24SNN and 24SNR, and the scores for 8SNN were significantly lower than for 24SNN, 24SNR, and 24SMR (all $p$\,$<$\,0.05).

We also found a significant main effect of the conditions on \emph{VD2}, $F(2.6,51.9)$\,=\,24.74, \textbf{\itshape{p}\,$<$\,0.001}, $\eta^2_p$\,=\,0.55.
Our post-hoc tests showed that the scores for 8DMR and 8SNN were both significantly lower than for 24SNN, 24SNR, and 24SMR (all $p$\,$<$\,0.05).

We further found a significant main effect of the conditions on \emph{VD3}, $F(2.6,52.8)$\,=\,15.43, \textbf{\itshape{p}\,$<$\,0.001}, $\eta^2_p$\,=\,0.44.
Our post-hoc tests showed that the scores for 8DMR were significantly lower than for 24SNN and 24SNR, and the scores for 8SNN were significantly lower than for 24SNN, 24SNR, and 24SMR (all $p$\,$<$\,0.05).

We only found a non-significant trend for an effect of the conditions on \emph{RE1}, $F(4,80)$\,=\,2.19, $p$\,=\,0.078, $\eta^2_p$\,=\,0.10.

Next, we further found a significant main effect of the conditions on \emph{RE2}, $F(2.6,51.5)$\,=\,3.52, \textbf{\itshape{p}\,=\,0.027}, $\eta^2_p$\,=\,0.15.
Our post-hoc tests showed that the scores for 8SNN were significantly lower than for 8DMR, and the scores for 8SNN were significantly lower than for 24SMR (all $p$\,$<$\,0.05).

Next, we further found a significant main effect of the conditions on \emph{ART}, $F(2.1,42.7)$\,=\,5.81, \textbf{\itshape{p}\,=\,0.005}, $\eta^2_p$\,=\,0.23.
Our post-hoc tests showed that the scores for 8DMR were significantly higher than for 24SNR and 24SMR, and the scores for 8SNN were significantly higher than for 24SMR (all $p$\,$<$\,0.05).

Additionally, we found a significant main effect of the three conditions on \emph{ROI}, $F(2,36)$\,=\,9.38, \textbf{\itshape{p}\,$<$\,0.001}, $\eta^2_p$\,=\,0.34.
Our post-hoc tests showed that the scores for 8DMR were significantly lower than for 24SNR and 24SMR (all $p$\,$<$\,0.05).

Last but not least, we found that the scores for 8DMR were significantly higher than for 24SMR for \emph{DU}, $F(1,19)$\,=\,29.19, \textbf{\itshape{p}\,$<$\,0.001}, $\eta^2_p$\,=\,0.61.

\subsubsection{Additional Items}
\label{sub:addition}

\paragraph{General Questions}
\label{sub:gq}
Participants indicated $M$\,=\,6.05 ($SD$\,=\,1.07) for \emph{GQ1}, $M$\,=\,6.67 ($SD$\,=\,1.24) for \emph{GQ2},  $M$\,=\,6.48 ($SD$\,=\,0.81) for \emph{GQ3}, and $M$\,=\,6.33 ($SD$\,=\,1.06) for \emph{GQ4}.
The results indicate a strong consensus regarding the importance of visual fidelity in VR telepresence. Participants reported that enhanced visual detail improved their ability to understand the scene and highlighted its value for practical telepresence applications. Respondents also emphasized the importance of perceiving fine visual details in real-world scenarios, such as remote education and meetings. In addition, many participants indicated that they were willing to tolerate minor visual artifacts in exchange for improved visual clarity.

\paragraph{Simulator Sickness}
We computed the difference between the pre and post SSQ scores, revealing a mean increase of 4.45 throughout the experiment.
Overall, this indicates only a minor increase in sickness symptoms for participants, in line with the short duration of this experiment.

\section{Discussion}
\label{sec:discussion}
We discuss the implications of our findings for multi-layer telepresence, focusing on background resolution, comparisons with super-resolution and native 8K video, and the remaining limitations of the dynamic update layer.

\subsection{Impact of Background Resolution}

Although 24K resolution is numerically much higher than 8K, it is still necessary to demonstrate that this increase translates into a meaningful improvement in user experience. Our results in Section~\ref{sub:ueq} and Section~\ref{sub:ipq} show that, compared with the 8K static background condition (8SNN), the 24K static background condition (24SNN) led to significantly better ratings in \emph{Overall Usability}, \emph{General Presence}, and \emph{Spatial Presence}. In addition, as indicated by the results of \emph{VD1}, \emph{VD2} and \emph{VD3} in Section~\ref{sub:siq}, textual and signage details in the 24K background were easier to read in VR, and the overall image was perceived as sharper. Responses to the general questions (Section~\ref{sub:gq}) further revealed a clear preference for high visual fidelity in VR telepresence.

Taken together, these findings suggest that increasing the background resolution from 8K to 24K provides clear benefits for perceived visual quality and for important aspects of user experience, including usability and presence.

\subsection{Comparison with Super-Resolution}

Real-time super-resolution for 8K 360-degree video remains difficult to achieve with current hardware and algorithmic constraints. Therefore, we compared our approach against a non-real-time state-of-the-art super-resolution method (Section~\ref{sub:sr}). The results indicate that, while super-resolution can enhance the appearance of an image, it cannot reliably recover scene details that are absent from the original input. As a result, it was unable to surpass our method in terms of perceived scene detail.

This suggests that, at present, super-resolution is not a sufficient solution to the limited resolution of native 8K 360-degree video, especially when the goal is to preserve fine visual details that are important for immersive telepresence viewing.

\subsection{Advantages over Native 8K Video}

Compared with native 8K video, our system without the dynamic update layer still provided a substantially better viewing experience in terms of overall sharpness and the clarity of fine scene details, as reflected in the results for \emph{VD1}, \emph{VD2}, and \emph{VD3} in Section~\ref{sub:siq} and Figure~\ref{fig:results_singleitem}. 
This suggests that our system configuration combining the UHR background and the ROI layer offers a clear advantage over a native 8K stream in the perception of scene detail, although this comparison also differs in background dynamics and the presence of a moving subject, so the effect cannot be attributed to resolution alone.

When the dynamic update layer was enabled, its resolution remained much lower than that of the UHR background layer. As a result, the system could not consistently outperform native 8K video on all aspects of visual quality, although it still showed an advantage in the perception of fine details (\emph{VD2}). This pattern is also reflected in participant feedback. For example, one participant wrote of the native 8K condition (8DMR), ``I noticed that the text was not readable,'' whereas the same participant commented on our system condition (24SMR), ``I was able to read the text properly.''

Furthermore, as indicated by the results for \emph{ROI} in Section~\ref{sub:siq}, enabling the ROI layer led to a clearly better viewing experience within the ROI region than native 8K video. Overall, these findings suggest that our layered representation can provide a superior viewing experience to native 8K video, particularly for fine details and ROI content, 
while the remaining limitations appear to be concentrated in the dynamic update layer.

\subsection{Artifacts and Misalignment} 

Visual artifacts were not unique to our system. Even native 8K video exhibited visible artifacts, as our 360-degree camera was composed of two fisheye lenses and the panoramic video was generated by stitching their views in real time. As a result, stitching seams and related artifacts could remain visible. Multiple participants reported similar issues in the questionnaire; for example, one participant noted that ``the card that had the number 13 and the maze image were cut off.'' This region corresponded to a stitching boundary between the two fisheye views. 

Compared with native video, our system introduced additional sources of artifacts. The background layer was generated from stitched high-resolution photographs, and the final representation further combined multiple visual layers. This multi-layer composition can lead to artifacts such as smearing and object misalignment. 

Object misalignment was most noticeable near the boundary between the ROI layer and the UHR background layer. Although both layers were derived from imagery captured by the PTZ camera, they were incorporated into the final scene in different ways. The ROI layer was displayed as a live video view, whereas the UHR background was constructed from PTZ photographs that were subsequently aligned to the 360-degree video representation. As a result, corresponding scene content could reflect slightly different viewpoints and geometric transformations, making perfect alignment between the live 4K layer and the background difficult to achieve. 

At the same time, increased visual fidelity made not only scene details but also rendering imperfections easier to perceive. In other words, the same improvement in sharpness that enhanced readability and fine-detail perception also made artifacts more visually salient. This interpretation is consistent with our results on \emph{ART} in Section~\ref{sub:siq}, where native 8K video received the most favorable evaluations, while the addition of more layers in our system tended to reduce these ratings. 

Taken together, these findings highlight an important trade-off in layered telepresence rendering: improving visual detail does not necessarily reduce perceived artifacts, and in some cases may make them more apparent. Future work should therefore focus not only on increasing scene fidelity, but also on improving cross-layer consistency and reducing visible transitions between layers.

\subsection{Limitations and Future Work}


Taken together, these observations highlight that the main remaining limitations of the current system lie in the dynamic update layer, which currently operates at 12.76 FPS with an end-to-end latency of approximately 1.41\,s. Two factors underlie this performance. The first is computational: foreground extraction requires resampling the incoming 8K panorama into cubemap faces, and this reprojection step dominates runtime. The second is visual: to preserve interactive performance, the system uses an efficient BackgroundMattingV2 configuration with a MobileNetV2 backbone, which yields relatively coarse extraction results. Since the extracted foreground is composited onto the UHR background layer, these imperfections can become particularly noticeable. Consistent with this, the results of \emph{DU} in Section~\ref{sub:siq} suggest that the moving subject did not always blend naturally with the background compared with native video.

We note that these figures partly reflect a deliberate design choice: the system targets the highest capture resolution available on current prosumer 360-degree cameras in order to explore the upper envelope of static-scene telepresence, and the source 8K live stream itself operates at only 30 FPS. Because the cost of reprojection and matting scales with capture resolution, the same dynamic update pipeline would achieve substantially higher frame rates at lower capture resolutions.
In practical terms, the current frame rate and latency are acceptable for observation-oriented scenarios such as remote tours and inspection of largely static environments---the settings our system primarily targets---whereas tightly coupled interactive telepresence, including synchronized audio and two-way communication, would demand lower latency and is left for future work.
Future improvements may therefore come from both stronger GPU resources and more advanced real-time matting or foreground extraction methods, which could improve both the frame rate and the visual consistency of the dynamic update layer.

On the hardware side, replacing the current copper strap mount with a rigid 3D-printed fixture could improve the mechanical stability and repeatability of the dual-camera alignment.

A further limitation concerns the interpretability of the user study. As the full system differs from native 8K streaming in several respects simultaneously (background resolution, background dynamics, and the layered compositing itself), its observed advantages reflect the combined effect of the system rather than any single component; only the adjacent-condition comparisons in Section~\ref{sec:experiment} support factor-level attribution. A 24K dynamic baseline that would fully disentangle these factors is not obtainable with current capture hardware, and a complete factorial evaluation remains future work.

Regarding our user study, another limitation to note is the gender imbalance of the participant sample (15 male, 6 female). Future evaluations should aim for a more balanced sample to improve the generalizability of the findings.

\section{Conclusion}
In this paper, we presented a multi-layer telepresence system that combines an ultra-high-resolution panoramic background, an ROI layer, and a dynamic update layer to support both global awareness and localized detail in VR telepresence. By exploiting the largely static nature of many telepresence scenes, the system separates offline background construction from real-time updates. 
Our evaluations showed that this design improves perceived sharpness and fine-detail readability compared with native 8K video and super-resolution-based alternatives, and that the ultra-high-resolution background improves usability and presence over its 8K counterpart, although artifacts and cross-layer inconsistencies remain, especially in the dynamic update layer.
Overall, these results suggest that multi-layer representations are a promising way to overcome the visual limitations of current 360-degree telepresence systems.


\bibliographystyle{abbrv-doi-hyperref}

\bibliography{template}

\begin{thebibliography}{10}

\bibitem{agustsson2017ntire}
E.~Agustsson and R.~Timofte.
\newblock Ntire 2017 challenge on single image super-resolution: Dataset and study.
\newblock In {\em 2017 IEEE Conference on Computer Vision and Pattern Recognition Workshops (CVPRW)}, pp. 1122--1131, 2017. \href{https://doi.org/10.1109/CVPRW.2017.150}
{doi: {{%
10\hspace{.1pt}\discretionary{.}{%
}{.}\hspace{.4pt}1109\discretionary{/}{%
}{/}CVPRW\hspace{.1pt}\discretionary{.}{%
}{.}\hspace{.4pt}2017\hspace{.1pt}\discretionary{.}{%
}{.}\hspace{.4pt}150}}}


\bibitem{ashraf2024resolution}
M.~Ashraf, A.~Chapiro, and R.~K. Mantiuk.
\newblock Resolution limit of the eye—how many pixels can we see?
\newblock {\em Nature Communications}, 16(1):9086, 2025. \href{https://doi.org/10.1038/s41467-025-64679-2}
{doi: {{%
10\hspace{.1pt}\discretionary{.}{%
}{.}\hspace{.4pt}1038\discretionary{/}{%
}{/}s41467\discretionary{%
}{-}{-}025\discretionary{%
}{-}{-}64679\discretionary{%
}{-}{-}2}}}


\bibitem{bharadwaj2009accommodative}
S.~R. Bharadwaj and T.~R. Candy.
\newblock Accommodative and vergence responses to conflicting blur and disparity stimuli during development.
\newblock {\em Journal of vision}, 9(11):4--4, 2009. \href{https://doi.org/10.1167/9.11.4}
{doi: {{%
10\hspace{.1pt}\discretionary{.}{%
}{.}\hspace{.4pt}1167\discretionary{/}{%
}{/}9\hspace{.1pt}\discretionary{.}{%
}{.}\hspace{.4pt}11\hspace{.1pt}\discretionary{.}{%
}{.}\hspace{.4pt}4}}}


\bibitem{chan2022investigating}
K.~C. Chan, S.~Zhou, X.~Xu, and C.~C. Loy.
\newblock Investigating tradeoffs in real-world video super-resolution.
\newblock In {\em 2022 IEEE/CVF Conference on Computer Vision and Pattern Recognition (CVPR)}, pp. 5952--5961, 2022. \href{https://doi.org/10.1109/CVPR52688.2022.00587}
{doi: {{%
10\hspace{.1pt}\discretionary{.}{%
}{.}\hspace{.4pt}1109\discretionary{/}{%
}{/}CVPR52688\hspace{.1pt}\discretionary{.}{%
}{.}\hspace{.4pt}2022\hspace{.1pt}\discretionary{.}{%
}{.}\hspace{.4pt}00587}}}


\bibitem{chen2023hat}
X.~Chen, X.~Wang, W.~Zhang, X.~Kong, Y.~Qiao, J.~Zhou et al.
\newblock Hat: Hybrid attention transformer for image restoration.
\newblock {\em IEEE Transactions on Pattern Analysis and Machine Intelligence}, 48(3):2676--2694, 2026. \href{https://doi.org/10.1109/TPAMI.2025.3628275}
{doi: {{%
10\hspace{.1pt}\discretionary{.}{%
}{.}\hspace{.4pt}1109\discretionary{/}{%
}{/}TPAMI\hspace{.1pt}\discretionary{.}{%
}{.}\hspace{.4pt}2025\hspace{.1pt}\discretionary{.}{%
}{.}\hspace{.4pt}3628275}}}


\bibitem{chi2025modular}
J.~Chi, C.~Neumann, C.~Cruz-Neira, and D.~Reiners.
\newblock A modular hybrid telepresence system integrating immersive 360-degree video with on-demand high-resolution imaging.
\newblock In {\em International Conference on Virtual Reality and Mixed Reality}, pp. 81--103. Springer, 2025. \href{https://doi.org/10.1007/978-3-032-03805-0_5}
{doi: {{%
10\hspace{.1pt}\discretionary{.}{%
}{.}\hspace{.4pt}1007\discretionary{/}{%
}{/}978\discretionary{%
}{-}{-}3\discretionary{%
}{-}{-}032\discretionary{%
}{-}{-}03805\discretionary{%
}{-}{-}0\_5}}}


\bibitem{elliott2020clinical}
D.~B. Elliott.
\newblock {\em Clinical procedures in primary eye care E-Book}.
\newblock Elsevier Health Sciences, 2020.

\bibitem{ref-gutenberg}
{Gutenberg}.
\newblock Free ebooks | project gutenberg.
\newblock \url{https://www.gutenberg.org/}, 2026.
\newblock Accessed: March 2026.

\bibitem{ha2020telepresence}
V.~K.~L. Ha, R.~Chai, and H.~T. Nguyen.
\newblock A telepresence wheelchair with 360-degree vision using webrtc.
\newblock {\em Applied Sciences}, 10(1), 2020. \href{https://doi.org/10.3390/app10010369}
{doi: {{%
10\hspace{.1pt}\discretionary{.}{%
}{.}\hspace{.4pt}3390\discretionary{/}{%
}{/}app10010369}}}


\bibitem{hilliard2024360u}
J.~Hilliard, A.~Hilton, and J.-Y. Guillemaut.
\newblock 360u-former: Hdr illumination estimation with panoramic adapted vision transformers.
\newblock In {\em European Conference on Computer Vision}, pp. 431--448. Springer, 2024. \href{https://doi.org/10.1007/978-3-031-91838-4_26}
{doi: {{%
10\hspace{.1pt}\discretionary{.}{%
}{.}\hspace{.4pt}1007\discretionary{/}{%
}{/}978\discretionary{%
}{-}{-}3\discretionary{%
}{-}{-}031\discretionary{%
}{-}{-}91838\discretionary{%
}{-}{-}4\_26}}}


\bibitem{hinderks2017design}
A.~Hinderks.
\newblock Design and evaluation of a short version of the user experience questionnaire (ueq-s).
\newblock {\em International Journal of Interactive Multimedia and Artificial Intelligence}, 2017. \href{https://doi.org/10.9781/ijimai.2017.09.001}
{doi: {{%
10\hspace{.1pt}\discretionary{.}{%
}{.}\hspace{.4pt}9781\discretionary{/}{%
}{/}ijimai\hspace{.1pt}\discretionary{.}{%
}{.}\hspace{.4pt}2017\hspace{.1pt}\discretionary{.}{%
}{.}\hspace{.4pt}09\hspace{.1pt}\discretionary{.}{%
}{.}\hspace{.4pt}001}}}


\bibitem{hsu2024drct}
C.-C. Hsu, C.-M. Lee, and Y.-S. Chou.
\newblock Drct: Saving image super-resolution away from information bottleneck.
\newblock In {\em 2024 IEEE/CVF Conference on Computer Vision and Pattern Recognition Workshops (CVPRW)}, pp. 6133--6142, 2024. \href{https://doi.org/10.1109/CVPRW63382.2024.00618}
{doi: {{%
10\hspace{.1pt}\discretionary{.}{%
}{.}\hspace{.4pt}1109\discretionary{/}{%
}{/}CVPRW63382\hspace{.1pt}\discretionary{.}{%
}{.}\hspace{.4pt}2024\hspace{.1pt}\discretionary{.}{%
}{.}\hspace{.4pt}00618}}}


\bibitem{ref-insta360pro2}
{Insta360}.
\newblock Insta360 pro 2.
\newblock \url{https://www.insta360.com/product/insta360-pro2}, 2025.
\newblock Accessed: September 2025.

\bibitem{ref-insta360x4}
{Insta360}.
\newblock Insta360 x4.
\newblock \url{https://www.insta360.com/product/insta360-x4}, 2025.
\newblock Accessed: September 2025.

\bibitem{ref-insta360x5}
{Insta360}.
\newblock Insta360 x5.
\newblock \url{https://www.insta360.com/product/insta360-x5}, 2025.
\newblock Accessed: September 2025.

\bibitem{kasahara2015jackin}
S.~Kasahara and J.~Rekimoto.
\newblock Jackin head: immersive visual telepresence system with omnidirectional wearable camera for remote collaboration.
\newblock In {\em Proceedings of the 21st ACM Symposium on Virtual Reality Software and Technology}, VRST '15, pp. 217–225. Association for Computing Machinery, New York, NY, USA, 2015. \href{https://doi.org/10.1145/2821592.2821608}
{doi: {{%
10\hspace{.1pt}\discretionary{.}{%
}{.}\hspace{.4pt}1145\discretionary{/}{%
}{/}2821592\hspace{.1pt}\discretionary{.}{%
}{.}\hspace{.4pt}2821608}}}


\bibitem{ke2022modnet}
Z.~Ke, J.~Sun, K.~Li, Q.~Yan, and R.~W. Lau.
\newblock Modnet: Real-time trimap-free portrait matting via objective decomposition.
\newblock In {\em Proceedings of the AAAI conference on artificial intelligence}, vol.~36, pp. 1140--1147, 2022. \href{https://doi.org/10.1609/aaai.v36i1.19999}
{doi: {{%
10\hspace{.1pt}\discretionary{.}{%
}{.}\hspace{.4pt}1609\discretionary{/}{%
}{/}aaai\hspace{.1pt}\discretionary{.}{%
}{.}\hspace{.4pt}v36i1\hspace{.1pt}\discretionary{.}{%
}{.}\hspace{.4pt}19999}}}


\bibitem{kennedy1993simulator}
R.~S. Kennedy, N.~E. Lane, K.~S. Berbaum, and M.~G. Lilienthal.
\newblock Simulator sickness questionnaire: An enhanced method for quantifying simulator sickness.
\newblock {\em The international journal of aviation psychology}, 3(3):203--220, 1993. \href{https://doi.org/10.1207/s15327108ijap0303_3}
{doi: {{%
10\hspace{.1pt}\discretionary{.}{%
}{.}\hspace{.4pt}1207\discretionary{/}{%
}{/}s15327108ijap0303\_3}}}


\bibitem{kramida2015resolving}
G.~Kramida.
\newblock Resolving the vergence-accommodation conflict in head-mounted displays.
\newblock {\em IEEE Transactions on Visualization and Computer Graphics}, 22(7):1912--1931, 2016. \href{https://doi.org/10.1109/TVCG.2015.2473855}
{doi: {{%
10\hspace{.1pt}\discretionary{.}{%
}{.}\hspace{.4pt}1109\discretionary{/}{%
}{/}TVCG\hspace{.1pt}\discretionary{.}{%
}{.}\hspace{.4pt}2015\hspace{.1pt}\discretionary{.}{%
}{.}\hspace{.4pt}2473855}}}


\bibitem{lawrence2024project}
J.~Lawrence, R.~Overbeck, T.~Prives, T.~Fortes, N.~Roth, and B.~Newman.
\newblock Project starline: A high-fidelity telepresence system.
\newblock In {\em ACM SIGGRAPH 2024 Emerging Technologies}, SIGGRAPH '24,  art. no. 16,  2 pp. Association for Computing Machinery, New York, NY, USA, 2024. \href{https://doi.org/10.1145/3641517.3664381}
{doi: {{%
10\hspace{.1pt}\discretionary{.}{%
}{.}\hspace{.4pt}1145\discretionary{/}{%
}{/}3641517\hspace{.1pt}\discretionary{.}{%
}{.}\hspace{.4pt}3664381}}}


\bibitem{li2023deep}
J.~Li, J.~Zhang, and D.~Tao.
\newblock Deep image matting: A comprehensive survey.
\newblock {\em arXiv preprint arXiv:2304.04672}, 2023. \href{https://doi.org/10.48550/arXiv.2304.04672}
{doi: {{%
10\hspace{.1pt}\discretionary{.}{%
}{.}\hspace{.4pt}48550\discretionary{/}{%
}{/}arXiv\hspace{.1pt}\discretionary{.}{%
}{.}\hspace{.4pt}2304\hspace{.1pt}\discretionary{.}{%
}{.}\hspace{.4pt}04672}}}


\bibitem{li2025mucvr}
W.~Li, Q.~Li, W.~Tian, J.~Gao, F.~Wu, J.~Liu et al.
\newblock Mucvr: Edge computing-enabled high-quality multi-user collaboration for interactive mvr.
\newblock {\em IEEE Transactions on Parallel and Distributed Systems}, 36(10):2058--2072, 2025. \href{https://doi.org/10.1109/TPDS.2025.3595801}
{doi: {{%
10\hspace{.1pt}\discretionary{.}{%
}{.}\hspace{.4pt}1109\discretionary{/}{%
}{/}TPDS\hspace{.1pt}\discretionary{.}{%
}{.}\hspace{.4pt}2025\hspace{.1pt}\discretionary{.}{%
}{.}\hspace{.4pt}3595801}}}


\bibitem{lin2021real}
S.~Lin, A.~Ryabtsev, S.~Sengupta, B.~Curless, S.~Seitz, and I.~Kemelmacher-Shlizerman.
\newblock Real-time high-resolution background matting.
\newblock In {\em 2021 IEEE/CVF Conference on Computer Vision and Pattern Recognition (CVPR)}, pp. 8758--8767, 2021. \href{https://doi.org/10.1109/CVPR46437.2021.00865}
{doi: {{%
10\hspace{.1pt}\discretionary{.}{%
}{.}\hspace{.4pt}1109\discretionary{/}{%
}{/}CVPR46437\hspace{.1pt}\discretionary{.}{%
}{.}\hspace{.4pt}2021\hspace{.1pt}\discretionary{.}{%
}{.}\hspace{.4pt}00865}}}


\bibitem{lindenberger2023lightglue}
P.~Lindenberger, P.-E. Sarlin, and M.~Pollefeys.
\newblock Lightglue: Local feature matching at light speed.
\newblock In {\em 2023 IEEE/CVF International Conference on Computer Vision (ICCV)}, pp. 17581--17592, 2023. \href{https://doi.org/10.1109/ICCV51070.2023.01616}
{doi: {{%
10\hspace{.1pt}\discretionary{.}{%
}{.}\hspace{.4pt}1109\discretionary{/}{%
}{/}ICCV51070\hspace{.1pt}\discretionary{.}{%
}{.}\hspace{.4pt}2023\hspace{.1pt}\discretionary{.}{%
}{.}\hspace{.4pt}01616}}}


\bibitem{lloyd2015practical}
C.~J. Lloyd, M.~Winterbottom, J.~Gaska, and L.~Williams.
\newblock A practical definition of eye-limited display system resolution.
\newblock In {\em Display Technologies and Applications for Defense, Security, and Avionics IX; and Head-and Helmet-Mounted Displays XX}, vol. 9470, pp. 107--115. SPIE, 2015. \href{https://doi.org/10.1117/12.2181077}
{doi: {{%
10\hspace{.1pt}\discretionary{.}{%
}{.}\hspace{.4pt}1117\discretionary{/}{%
}{/}12\hspace{.1pt}\discretionary{.}{%
}{.}\hspace{.4pt}2181077}}}


\bibitem{lowe2004distinctive}
D.~G. Lowe.
\newblock Distinctive image features from scale-invariant keypoints.
\newblock {\em International journal of computer vision}, 60(2):91--110, 2004. \href{https://doi.org/10.1023/B:VISI.0000029664.99615.94}
{doi: {{%
10\hspace{.1pt}\discretionary{.}{%
}{.}\hspace{.4pt}1023\discretionary{/}{%
}{/}B\discretionary{:}{%
}{:}VISI\hspace{.1pt}\discretionary{.}{%
}{.}\hspace{.4pt}0000029664\hspace{.1pt}\discretionary{.}{%
}{.}\hspace{.4pt}99615\hspace{.1pt}\discretionary{.}{%
}{.}\hspace{.4pt}94}}}


\bibitem{luo2024comparative}
Q.~Luo, J.~Zhang, Y.~Xie, X.~Huang, and T.~Han.
\newblock Comparative analysis of advanced feature matching algorithms in challenging high spatial resolution optical satellite stereo scenarios.
\newblock In {\em IGARSS 2024 - 2024 IEEE International Geoscience and Remote Sensing Symposium}, pp. 2645--2649, 2024. \href{https://doi.org/10.1109/IGARSS53475.2024.10641727}
{doi: {{%
10\hspace{.1pt}\discretionary{.}{%
}{.}\hspace{.4pt}1109\discretionary{/}{%
}{/}IGARSS53475\hspace{.1pt}\discretionary{.}{%
}{.}\hspace{.4pt}2024\hspace{.1pt}\discretionary{.}{%
}{.}\hspace{.4pt}10641727}}}


\bibitem{ref-metaquest3}
{Meta}.
\newblock Meta quest 3.
\newblock \url{https://www.meta.com/quest/quest-3/}, 2025.
\newblock Accessed: September 2025.

\bibitem{minsky1980telepresence}
M.~Minsky.
\newblock Telepresence.
\newblock 1980.

\bibitem{narciso2019immersive}
D.~Narciso, M.~Bessa, M.~Melo, A.~Coelho, and J.~Vasconcelos-Raposo.
\newblock Immersive 360∘ video user experience: impact of different variables in the sense of presence and cybersickness.
\newblock {\em Universal Access in the Information Society}, 18(1):77--87, 2019. \href{https://doi.org/10.1007/s10209-017-0581-5}
{doi: {{%
10\hspace{.1pt}\discretionary{.}{%
}{.}\hspace{.4pt}1007\discretionary{/}{%
}{/}s10209\discretionary{%
}{-}{-}017\discretionary{%
}{-}{-}0581\discretionary{%
}{-}{-}5}}}


\bibitem{nassani2021showmearound}
A.~Nassani, L.~Zhang, H.~Bai, and M.~Billinghurst.
\newblock Showmearound: Giving virtual tours using live 360 video.
\newblock In {\em Extended Abstracts of the 2021 CHI Conference on Human Factors in Computing Systems}, CHI EA '21,  art. no. 168,  4 pp. Association for Computing Machinery, New York, NY, USA, 2021. \href{https://doi.org/10.1145/3411763.3451555}
{doi: {{%
10\hspace{.1pt}\discretionary{.}{%
}{.}\hspace{.4pt}1145\discretionary{/}{%
}{/}3411763\hspace{.1pt}\discretionary{.}{%
}{.}\hspace{.4pt}3451555}}}


\bibitem{nguyen2020screen}
J.~Nguyen, C.~Smith, Z.~Magoz, and J.~Sears.
\newblock {Screen door effect reduction using mechanical shifting for virtual reality displays}.
\newblock In B.~C. Kress and C.~Peroz, eds., {\em Optical Architectures for Displays and Sensing in Augmented, Virtual, and Mixed Reality (AR, VR, MR)}, vol. 11310, p. 113100P. International Society for Optics and Photonics, SPIE, 2020. \href{https://doi.org/10.1117/12.2544479}
{doi: {{%
10\hspace{.1pt}\discretionary{.}{%
}{.}\hspace{.4pt}1117\discretionary{/}{%
}{/}12\hspace{.1pt}\discretionary{.}{%
}{.}\hspace{.4pt}2544479}}}


\bibitem{cupy_learningsys2017}
R.~Okuta, Y.~Unno, D.~Nishino, S.~Hido, and C.~Loomis.
\newblock Cupy: A numpy-compatible library for nvidia gpu calculations.
\newblock In {\em Proceedings of Workshop on Machine Learning Systems (LearningSys) in The Thirty-first Annual Conference on Neural Information Processing Systems (NIPS)}, 2017.

\bibitem{pejsa2016room2room}
T.~Pejsa, J.~Kantor, H.~Benko, E.~Ofek, and A.~Wilson.
\newblock Room2room: Enabling life-size telepresence in a projected augmented reality environment.
\newblock In {\em Proceedings of the 19th ACM Conference on Computer-Supported Cooperative Work \& Social Computing}, CSCW '16, pp. 1716–1725. Association for Computing Machinery, New York, NY, USA, 2016. \href{https://doi.org/10.1145/2818048.2819965}
{doi: {{%
10\hspace{.1pt}\discretionary{.}{%
}{.}\hspace{.4pt}1145\discretionary{/}{%
}{/}2818048\hspace{.1pt}\discretionary{.}{%
}{.}\hspace{.4pt}2819965}}}


\bibitem{perez2003poisson}
P.~P\'{e}rez, M.~Gangnet, and A.~Blake.
\newblock {\em Poisson Image Editing}.
\newblock Association for Computing Machinery, New York, NY, USA, 1 ed., 2023. \href{https://doi.org/10.1145/3596711.3596772}
{doi: {{%
10\hspace{.1pt}\discretionary{.}{%
}{.}\hspace{.4pt}1145\discretionary{/}{%
}{/}3596711\hspace{.1pt}\discretionary{.}{%
}{.}\hspace{.4pt}3596772}}}


\bibitem{ref-pimax}
{Pimax}.
\newblock Pimax crystal super.
\newblock \url{https://pimax.com/pages/pimax-crystal-super}, 2025.
\newblock Accessed: September 2025.

\bibitem{reinhard2002color}
E.~Reinhard, M.~Adhikhmin, B.~Gooch, and P.~Shirley.
\newblock Color transfer between images.
\newblock {\em IEEE Computer Graphics and Applications}, 21(5):34--41, 2001. \href{https://doi.org/10.1109/38.946629}
{doi: {{%
10\hspace{.1pt}\discretionary{.}{%
}{.}\hspace{.4pt}1109\discretionary{/}{%
}{/}38\hspace{.1pt}\discretionary{.}{%
}{.}\hspace{.4pt}946629}}}


\bibitem{schubert2001experience}
T.~Schubert, F.~Friedmann, and H.~Regenbrecht.
\newblock The experience of presence: Factor analytic insights.
\newblock {\em Presence: Teleoper. Virtual Environ.}, 10(3):266–281, June 2001. \href{https://doi.org/10.1162/105474601300343603}
{doi: {{%
10\hspace{.1pt}\discretionary{.}{%
}{.}\hspace{.4pt}1162\discretionary{/}{%
}{/}105474601300343603}}}


\bibitem{sengupta2020background}
S.~Sengupta, V.~Jayaram, B.~Curless, S.~M. Seitz, and I.~Kemelmacher-Shlizerman.
\newblock Background matting: The world is your green screen.
\newblock In {\em 2020 IEEE/CVF Conference on Computer Vision and Pattern Recognition (CVPR)}, pp. 2288--2297, 2020. \href{https://doi.org/10.1109/CVPR42600.2020.00236}
{doi: {{%
10\hspace{.1pt}\discretionary{.}{%
}{.}\hspace{.4pt}1109\discretionary{/}{%
}{/}CVPR42600\hspace{.1pt}\discretionary{.}{%
}{.}\hspace{.4pt}2020\hspace{.1pt}\discretionary{.}{%
}{.}\hspace{.4pt}00236}}}


\bibitem{shen2014viewing}
C.-T. Shen, H.-H. Liu, M.-H. Yang, Y.-P. Hung, and S.-C. Pei.
\newblock Viewing-distance aware super-resolution for high-definition display.
\newblock {\em IEEE Transactions on Image Processing}, 24(1):403--418, 2015. \href{https://doi.org/10.1109/TIP.2014.2375639}
{doi: {{%
10\hspace{.1pt}\discretionary{.}{%
}{.}\hspace{.4pt}1109\discretionary{/}{%
}{/}TIP\hspace{.1pt}\discretionary{.}{%
}{.}\hspace{.4pt}2014\hspace{.1pt}\discretionary{.}{%
}{.}\hspace{.4pt}2375639}}}


\bibitem{shen2022panoformer}
Z.~Shen, C.~Lin, K.~Liao, L.~Nie, Z.~Zheng, and Y.~Zhao.
\newblock Panoformer: panorama transformer for indoor 360∘ depth estimation.
\newblock In {\em European Conference on Computer Vision}, pp. 195--211. Springer, 2022. \href{https://doi.org/10.1007/978-3-031-19769-7_12}
{doi: {{%
10\hspace{.1pt}\discretionary{.}{%
}{.}\hspace{.4pt}1007\discretionary{/}{%
}{/}978\discretionary{%
}{-}{-}3\discretionary{%
}{-}{-}031\discretionary{%
}{-}{-}19769\discretionary{%
}{-}{-}7\_12}}}


\bibitem{ref-spout}
{Spout}.
\newblock Ultra-fast, realtime video routing for windows.
\newblock \url{https://spout.zeal.co/}, 2026.
\newblock Accessed: July 2026.

\bibitem{sun2021loftr}
J.~Sun, Z.~Shen, Y.~Wang, H.~Bao, and X.~Zhou.
\newblock Loftr: Detector-free local feature matching with transformers.
\newblock In {\em 2021 IEEE/CVF Conference on Computer Vision and Pattern Recognition (CVPR)}, pp. 8918--8927, 2021. \href{https://doi.org/10.1109/CVPR46437.2021.00881}
{doi: {{%
10\hspace{.1pt}\discretionary{.}{%
}{.}\hspace{.4pt}1109\discretionary{/}{%
}{/}CVPR46437\hspace{.1pt}\discretionary{.}{%
}{.}\hspace{.4pt}2021\hspace{.1pt}\discretionary{.}{%
}{.}\hspace{.4pt}00881}}}


\bibitem{tyszkiewicz2020disk}
M.~J. Tyszkiewicz, P.~Fua, and E.~Trulls.
\newblock Disk: learning local features with policy gradient.
\newblock In {\em Proceedings of the 34th International Conference on Neural Information Processing Systems}, NIPS '20,  art. no. 1195,  12 pp. Curran Associates Inc., Red Hook, NY, USA, 2020.

\bibitem{ref-varjo}
{Varjo}.
\newblock Varjo xr-3 - the first true mixed reality headset | varjo.
\newblock \url{https://varjo.com/products/varjo-xr-3}, 2026.
\newblock Accessed: March 2026.

\bibitem{verma2019testing}
J.~P. Verma and A.-S.~G. Abdel-Salam.
\newblock {\em Testing statistical assumptions in research}.
\newblock John Wiley \& Sons, 2019.

\bibitem{ref-vrcompare-varjo}
{VRCompare}.
\newblock Varjo xr-3: Full specification - vrcompare.
\newblock \url{https://vr-compare.com/headset/varjoxr-3}, 2026.
\newblock Accessed: March 2026.

\bibitem{wang2021real}
X.~Wang, L.~Xie, C.~Dong, and Y.~Shan.
\newblock Real-esrgan: Training real-world blind super-resolution with pure synthetic data.
\newblock In {\em 2021 IEEE/CVF International Conference on Computer Vision Workshops (ICCVW)}, pp. 1905--1914, 2021. \href{https://doi.org/10.1109/ICCVW54120.2021.00217}
{doi: {{%
10\hspace{.1pt}\discretionary{.}{%
}{.}\hspace{.4pt}1109\discretionary{/}{%
}{/}ICCVW54120\hspace{.1pt}\discretionary{.}{%
}{.}\hspace{.4pt}2021\hspace{.1pt}\discretionary{.}{%
}{.}\hspace{.4pt}00217}}}


\bibitem{xu2018gaze}
Y.~Xu, Y.~Dong, J.~Wu, Z.~Sun, Z.~Shi, J.~Yu et al.
\newblock Gaze prediction in dynamic 360° immersive videos.
\newblock In {\em 2018 IEEE/CVF Conference on Computer Vision and Pattern Recognition}, pp. 5333--5342, 2018. \href{https://doi.org/10.1109/CVPR.2018.00559}
{doi: {{%
10\hspace{.1pt}\discretionary{.}{%
}{.}\hspace{.4pt}1109\discretionary{/}{%
}{/}CVPR\hspace{.1pt}\discretionary{.}{%
}{.}\hspace{.4pt}2018\hspace{.1pt}\discretionary{.}{%
}{.}\hspace{.4pt}00559}}}


\bibitem{yan2024panovos}
S.~Yan, X.~Xu, R.~Zhang, L.~Hong, W.~Chen, W.~Zhang et al.
\newblock Panovos: Bridging non-panoramic and panoramic views with transformer for video segmentation.
\newblock In A.~Leonardis, E.~Ricci, S.~Roth, O.~Russakovsky, T.~Sattler, and G.~Varol, eds., {\em Computer Vision -- ECCV 2024}, pp. 346--365. Springer Nature Switzerland, Cham, 2025. \href{https://doi.org/10.1007/978-3-031-72673-6_19}
{doi: {{%
10\hspace{.1pt}\discretionary{.}{%
}{.}\hspace{.4pt}1007\discretionary{/}{%
}{/}978\discretionary{%
}{-}{-}3\discretionary{%
}{-}{-}031\discretionary{%
}{-}{-}72673\discretionary{%
}{-}{-}6\_19}}}


\bibitem{ref-zeromq}
{ZeroMQ Project}.
\newblock Zeromq: An open-source universal messaging library.
\newblock \url{https://zeromq.org/}, 2026.
\newblock Accessed: July 2026.

\bibitem{zhang2013viewport}
C.~Zhang, Q.~Cai, P.~A. Chou, Z.~Zhang, and R.~Martin-Brualla.
\newblock Viewport: A distributed, immersive teleconferencing system with infrared dot pattern.
\newblock {\em IEEE MultiMedia}, 20(1):17--27, 2013. \href{https://doi.org/10.1109/MMUL.2013.12}
{doi: {{%
10\hspace{.1pt}\discretionary{.}{%
}{.}\hspace{.4pt}1109\discretionary{/}{%
}{/}MMUL\hspace{.1pt}\discretionary{.}{%
}{.}\hspace{.4pt}2013\hspace{.1pt}\discretionary{.}{%
}{.}\hspace{.4pt}12}}}


\bibitem{zhang2018detection}
J.~Zhang, E.~Langbehn, D.~Krupke, N.~Katzakis, and F.~Steinicke.
\newblock Detection thresholds for rotation and translation gains in 360° video-based telepresence systems.
\newblock {\em IEEE Transactions on Visualization and Computer Graphics}, 24(4):1671--1680, 2018. \href{https://doi.org/10.1109/TVCG.2018.2793679}
{doi: {{%
10\hspace{.1pt}\discretionary{.}{%
}{.}\hspace{.4pt}1109\discretionary{/}{%
}{/}TVCG\hspace{.1pt}\discretionary{.}{%
}{.}\hspace{.4pt}2018\hspace{.1pt}\discretionary{.}{%
}{.}\hspace{.4pt}2793679}}}


\bibitem{zhang2025leader360v}
W.~Zhang, D.~Xiao, A.~Dai, Y.~Liu, T.~Pan, S.~Wen et al.
\newblock Leader360v: The large-scale, real-world 360 video dataset for multi-task learning in diverse environment, 2025. \href{https://doi.org/10.48550/arXiv.2506.14271}
{doi: {{%
10\hspace{.1pt}\discretionary{.}{%
}{.}\hspace{.4pt}48550\discretionary{/}{%
}{/}arXiv\hspace{.1pt}\discretionary{.}{%
}{.}\hspace{.4pt}2506\hspace{.1pt}\discretionary{.}{%
}{.}\hspace{.4pt}14271}}}


\bibitem{zhang2022fast}
Y.~Zhang, F.-L. Zhang, Z.~Zhu, L.~Wang, and Y.~Jin.
\newblock Fast edit propagation for 360 degree panoramas using function interpolation.
\newblock {\em IEEE Access}, 10:43882--43894, 2022. \href{https://doi.org/10.1109/ACCESS.2022.3168665}
{doi: {{%
10\hspace{.1pt}\discretionary{.}{%
}{.}\hspace{.4pt}1109\discretionary{/}{%
}{/}ACCESS\hspace{.1pt}\discretionary{.}{%
}{.}\hspace{.4pt}2022\hspace{.1pt}\discretionary{.}{%
}{.}\hspace{.4pt}3168665}}}


\bibitem{zhong2025omnisam}
D.~Zhong, X.~Zheng, C.~Liao, Y.~Lyu, J.~Chen, S.~Wu et al.
\newblock Omnisam: Omnidirectional segment anything model for uda in panoramic semantic segmentation.
\newblock In {\em 2025 IEEE/CVF International Conference on Computer Vision (ICCV)}, pp. 23892--23901, 2025. \href{https://doi.org/10.1109/ICCV51701.2025.02215}
{doi: {{%
10\hspace{.1pt}\discretionary{.}{%
}{.}\hspace{.4pt}1109\discretionary{/}{%
}{/}ICCV51701\hspace{.1pt}\discretionary{.}{%
}{.}\hspace{.4pt}2025\hspace{.1pt}\discretionary{.}{%
}{.}\hspace{.4pt}02215}}}


\end{thebibliography}

\end{document}